%% file: main.tex
\pdfoutput=1

\documentclass[a4paper,fleqn]{cas-dc}

\usepackage{amsmath,amssymb,booktabs,graphicx,rotating}

\usepackage[caption=false]{subfig}
\graphicspath{{./figures/}}

\usepackage[per-mode=symbol]{siunitx}
\usepackage{algorithm}
\usepackage{algpseudocode}
\floatstyle{ruled}\restylefloat{algorithm}

\usepackage{microtype}

\usepackage{xcolor}
\definecolor{cVehA}{HTML}{00509B}
\definecolor{cVehB}{HTML}{E77B29}
\definecolor{cVehC}{HTML}{006400}
\definecolor{cBaseline}{HTML}{A694BE}
\definecolor{cAugmented}{HTML}{6A4C93}
\definecolor{cSDA}{HTML}{B23A48}
\definecolor{cTimeShift}{HTML}{C49A00}
\definecolor{cIMUBias}{HTML}{006F89}

\newcommand{\vy}{v_y}

\newcommand{\rot}[1]{\rotatebox{90}{#1}}
\newcommand{\tyes}{\checkmark}
\newcommand{\tpart}{($\sim$)}
\newcommand{\tno}{--}

\newcommand{\mm}[1]{\boldsymbol{#1}}
\newcommand{\mc}[1]{\mathcal{#1}}

\usepackage[numbers,sort&compress]{natbib}

\makeatletter
\AtEndDocument{\clearpage\immediate\write\@auxout{\string\csxdef{lastpage}{\number\numexpr\value{page}-1\relax}}}
\makeatother
\renewcommand{\printorcid}{}

\ExplSyntaxOn
\cs_set:Npn \__first_footerline:
{ \group_begin: \small \sffamily \__short_authors: :~
  { \rmfamily \itshape Preprint~ submitted~ to~ Control~ Engineering~ Practice } \group_end: }
\ExplSyntaxOff

\begin{document}

\let\WriteBookmarks\relax
\shorttitle{Cross-Vehicle Transferability of Lateral Velocity Estimation}
\shortauthors{H. Sch\"afke et al.}

\title[mode=title]{Cross-Vehicle Transferability of Lateral Velocity Estimation on Production Vehicles through Input Selection and Data Augmentation}

\author[imes]{Hendrik Sch\"afke}
\cormark[1]
\ead{schaefke@imes.uni-hannover.de}
\credit{Conceptualization, Data curation, Formal analysis, Investigation, Methodology, Project administration, Resources, Software, Validation, Visualization, Writing -- original draft, Writing -- review \& editing}
\author[imes]{Simon F. G. Ehlers}
\credit{Conceptualization, Funding acquisition, Methodology, Resources, Supervision, Writing -- review \& editing}
\author[imes,l3s]{Thomas Seel}
\credit{Conceptualization, Funding acquisition, Supervision, Writing -- review \& editing}
\author[iav]{Askar Vagapov}
\credit{Investigation, Resources, Writing -- review \& editing}
\author[iav]{Christoph Schweers}
\fnmark[1]
\credit{Investigation, Resources, Supervision}
\author[iav]{Ahmed Trabelsi}
\credit{Funding acquisition, Resources, Supervision}
\author[imes]{Daniel O. M. Weber}
\credit{Methodology, Supervision, Writing -- review \& editing}

\affiliation[imes]{organization={Institute of Mechatronic Systems, Leibniz University Hannover},
  postcode={30823}, city={Garbsen}, country={Germany}}
\affiliation[iav]{organization={IAV GmbH},
  postcode={38518}, city={Gifhorn}, country={Germany}}
\affiliation[l3s]{organization={L3S Research Center, Leibniz University Hannover},
  postcode={30167}, city={Hannover}, country={Germany}}

\cortext[1]{Corresponding author.}
\fntext[1]{C.~Schweers was affiliated with IAV GmbH when this work was performed.}

\begin{abstract} 
Lateral velocity is safety-critical for vehicle-dynamics control but cannot be measured with production sensors.
Model-based estimators such as Kalman filters are established, whereas data-driven estimators are more accurate within the training regime but suspected to lose accuracy beyond it.
Their transfer to a structurally different vehicle has rarely been tested.
We record real-world test drives of three series-production vehicles, train a neural-network estimator on production CAN signals of a compact car, and evaluate it zero-shot on an SUV and a van against a GNSS-aided inertial reference.
On the unseen vehicles, the network's error exceeds that on the training vehicle by an order of magnitude, a gap not quantified on production vehicles before.
A systematic comparison of input sets from ten production CAN signals identifies a five-signal set that transfers most robustly.
To narrow this gap, we introduce three physically motivated augmentations of training data.
Two improve generalization, namely exploiting the vehicle's left-right symmetry and emulating accelerometer bias.
Combined, they reduce the error on the validation maneuvers of the SUV and van by \SI{13}{\percent} and \SI{15}{\percent}.
Contrary to common expectation, on the SUV the data-driven estimator trained with both augmentations matches a model-based estimator tuned on the same training data.
\end{abstract}


\maketitle

\section{Introduction}
\label{sec:intro}
Advanced driver-assistance systems and automated driving functions rely on accurate knowledge of the vehicle's motion states~\citep{Jin2019}.
Among these states, the lateral velocity $\vy$ is particularly relevant for lateral control functions such as electronic stability control, torque vectoring, and evasive steering assistance.
Unlike the yaw rate or the wheel speeds, however, $\vy$ cannot be measured with production sensors.
Direct measurement requires optical sensors or high-precision GNSS-aided inertial systems, whose cost prevents series deployment.
The lateral velocity must therefore be estimated from the signals a production vehicle already provides.
As surveyed in~\citep{Jin2019}, two estimator families have emerged for this task.
Model-based estimators typically combine a physical vehicle model with a Kalman filter or a nonlinear observer~\citep{Puscul2024}, whereas data-driven estimators, in this work neural networks, learn the mapping from measured signals to $\vy$ directly.
Data-driven estimators have recently achieved strong accuracy~\citep{Giuliacci2023, Srinivasan2020, NapolitanoDellAnnunziata2024}, and where the two families have been compared directly, the data-driven estimator reached the higher average in-domain accuracy~\citep{James2020, Ziaukas2021, Kim2020}.\looseness=-1

Data-driven estimators are, however, expected to generalize poorly beyond their training distribution, both to unseen maneuvers and driving conditions~\citep{Kim2020} and, in particular, to unseen vehicles, because a neural network learns the specific dynamics of its training vehicle~\citep{Graeber2019, Liu2021}.
In practice, every estimator is designed, trained, and validated for one vehicle at a time.
Each additional vehicle platform or derivative requires repeating the reference measurements, the training, and the validation.
An estimator that transfers to further vehicles without target-vehicle data would, in the long run, turn this recurring effort into a one-time effort.
Whether or not full transfer is achieved, understanding how the estimation degrades across vehicles shows which parts of a data-driven estimator are vehicle-specific and how sensitive it is to changes in the vehicle parameters.
It also offers insight into how strongly the estimator depends on the particular dataset it was trained on.
This understanding is a step toward a potential vehicle-independent estimator.\looseness=-1

Whether data-driven estimators actually fail on unseen vehicles has rarely been tested directly.
Most data-driven work targets either $\vy$ or the sideslip angle $\beta = \arctan(\vy/v_x)$.
Given the longitudinal velocity $v_x$, either quantity follows from the other, and prior work applies the same estimator designs to both.
The following overview therefore includes estimators of $\beta$ as relevant prior work for $\vy$ estimation.
These estimators span pure neural networks and hybrids with a vehicle model or a filter, yet almost all of them are evaluated on the single vehicle they were trained on~\citep{Melzi2011, Ghosh2018, Graeber2019, Bonfitto2020, Gonzalez2020, Hermansdorfer2020, Kim2020, Srinivasan2020, Ziaukas2021, Giuliacci2023, Bertipaglia2024, Chen2024, Hwang2024, NapolitanoDellAnnunziata2024, Tristano2024PINN, Kalyanasundaram2025}.

Of the 18 data-driven lateral-velocity studies considered here (Table~\ref{tab:novelty}), only one aims at improving the transfer from a training vehicle to a different, unseen vehicle~\citep{DaLio2023}.
It embeds kinematic structure into the neural network and evaluates it on two race cars of the 1960s.
Trained without data of the second car, the structured neural network shows only a slight loss of accuracy on that car, whereas a black-box neural network collapses.
The only model-based estimator applied to both cars is a kinematic observer, which contains no vehicle parameters.
None of the studies considered here tests an estimator based on a dynamic vehicle model on a vehicle other than the one it was tuned for.
A second study on the same public dataset of the two race cars observes a significant error increase on the unseen vehicle and focuses on detecting the affected predictions by uncertainty quantification~\citep{Ayyad2021}.
Cross-vehicle evaluation requires reference-grade measurements on several vehicles, which are time-consuming to collect, and open datasets covering the signals of the production controller area network (CAN) barely exist.
For short-horizon yaw-rate forecasting, \citet{Kontos2025} train a neural network on a mid-size SUV, evaluate it without retraining on a smaller hatchback and two larger SUVs, and publish their data openly.
Forecasting predicts the future course of a signal the vehicle already measures, whereas estimating $\vy$ reconstructs a state that no production sensor provides.
Their published dataset accordingly contains no reference measurement and cannot support cross-vehicle $\vy$ estimation.
More broadly, the cross-vehicle transfer of data-driven vehicle-state estimators has, to the best of our knowledge, not been systematically examined on production passenger vehicles for any state.
This work addresses this gap for the lateral velocity, for which recent state-of-the-art estimators explicitly defer transfer to future work~\citep{Kalyanasundaram2025}.
We attribute this transfer problem to two sources.
The \emph{vehicle gap} arises from the differing physical parameters of the vehicles, and the \emph{measurement gap} arises from the measurement chain and the recording conditions.
Because each vehicle contributes a single recording campaign, both gaps occur together in every transfer of this study and cannot be separated with these data.
Section~\ref{sec:methods} details them.

\begin{table}[pos=t]
\centering
\footnotesize
\setlength{\tabcolsep}{4pt}
\renewcommand{\arraystretch}{1.15}
\caption{Positioning against prior neural-network lateral-velocity studies.
Of the 18 surveyed studies, the table lists those that address at least one of cross-vehicle evaluation, data augmentation, and systematic input selection.
The remaining surveyed studies address none of the three.
\citet{James2020} model the longitudinal dynamics and \citet{Kontos2025} forecast the yaw rate, and both are listed in addition because they address the cross-vehicle question.\looseness=-1}
\label{tab:novelty}
\begin{tabular*}{\columnwidth}{@{\extracolsep{\fill}}llccccc@{}}
\toprule
Study & Target &
\rot{Production vehicle~} &
\rot{Cross-vehicle~} &
\rot{Production CAN only~} &
\rot{Data augmentation~} &
\rot{Input-set selection~} \\
\midrule
\citet{Melzi2011}        & $\beta$            & \tyes  & \tno   & \tyes  & \tyes & \tno \\
\citet{Hwang2024}        & $\vy$              & \tno   & \tno   & \tyes  & \tno  & \tyes \\
\citet{Tristano2024PINN} & $\beta$  & \tpart & \tno   & \tyes  & \tno  & \tyes \\
\citet{Ayyad2021}        & $v_x$, $\vy$       & \tpart & \tpart & \tno   & \tno  & \tno \\
\citet{DaLio2023}        & $\vy$              & \tpart & \tyes  & \tyes  & \tno  & \tno \\
\citet{James2020}        & $v_x$              & \tyes  & \tpart & \tpart & \tno  & \tno \\
\citet{Kontos2025}       & $\dot\psi$ (pred.) & \tyes  & \tyes  & \tyes  & \tno  & \tno \\
\midrule
\textbf{This work} & $\vy$ & \textbf{\tyes} & \textbf{\tyes} & \textbf{\tyes} & \textbf{\tyes} & \textbf{\tyes} \\
\bottomrule
\end{tabular*}

\vspace{2pt}
{\footnotesize\raggedright
\tyes~addressed,\quad \tpart~partial or with caveats,\quad \tno~not addressed.\\
Production vehicle: \tpart~real non-production vehicle, \tno~simulation.\\
Targets: $\beta$~sideslip angle, $\vy$~lateral velocity, $\dot\psi$~(pred.)~short-horizon yaw-rate prediction, $v_x$~longitudinal velocity.\par}
\end{table}

Without data of the unseen vehicle, the transfer could be improved through the design of the estimator, its training data, or its input set.
Within a single vehicle, prior work improves the generalization to unseen maneuvers and road conditions by modifying the estimator itself, for example by embedding physical vehicle models into the neural network or coupling it with a model-based filter.
\citet{Graeber2019} append the sideslip rate of a kinematic single-track model to the inputs of a recurrent neural network, and \citet{Chen2024} let a recurrent neural network learn the residual dynamics relative to a linear single-track model.
\citet{Kim2020} feed the sideslip estimate of a neural network into a Kalman filter as a virtual measurement weighted by its reported uncertainty.
\citet{Bertipaglia2024} extend this coupling and train the neural network end to end through the filter.
\citet{Bonfitto2020} train separate neural networks for dry, wet, and icy roads and switch between them with a road-condition classifier.
\citet{Kalyanasundaram2025} fuse a transformer-based estimator with a kinematic model and a Kalman filter on a linear single-track model, weighting each branch by its reported uncertainty.
None of these designs has been tested on another vehicle, and those that embed a dynamic vehicle model need the parameters of each new vehicle.\looseness=-1

Data augmentation instead modifies the training data and leaves the estimator unchanged, and it is an established means of improving the generalization of time-series models~\citep{Wen2021, Stroebel2024}.
In vehicle-dynamics estimation, however, it has hardly been used.
To our knowledge, only \citet{Melzi2011} apply it, adding white measurement noise to simulated data for robustness.
\citet{Graeber2019} propose simulating sensor noise, offsets, and failures during training as future work.\looseness=-1

Physically motivated augmentations derive the transformation from a known property of the system or its measurement chain, such as a symmetry, a delay, or an offset, so that every generated sample remains a plausible measurement.
In the terminology of physics-informed machine learning, such augmentations introduce an observational bias, which complements the inductive bias built into an architecture~\citep{Karniadakis2021}.
In structural health monitoring, for example, \citet{Li2025} mirror the sensor layout of a simulated symmetric bridge to nearly double its damage cases.
Road vehicles are approximately mirror-symmetric about their longitudinal plane, and the delays and offsets of their sensors differ between vehicles and recording campaigns.
Physically motivated augmentation, which could therefore improve the transfer without data of the unseen vehicle, has not been applied to lateral-velocity or sideslip estimation.

Although the input set is a crucial factor in the accuracy of a data-driven estimator, almost every study considered here specifies a single set, and only two examine this choice.
\citet{Hwang2024} rank 14 candidate signals of a simulated vehicle dataset by correlation and mutual information and then compare three input sets.
\citet{Tristano2024PINN} compare subsets of a five-signal pool on public data and recommend a two-signal set that is at least as accurate as the full pool.
Both judge the input set by its in-domain accuracy and cover only part of the production CAN signals.
How the choice among these signals affects the transfer to an unseen vehicle has therefore not been examined.

This work examines these open questions on three structurally different production vehicles.
Vehicle~A (compact car) serves as the single training vehicle, and Vehicle~B (SUV) and Vehicle~C (van) are evaluated zero-shot.
Zero-shot means that no data of the unseen vehicle enter the training.
The input set and the three augmentations act on the data and leave the estimator itself unchanged.
We demonstrate them on a gated recurrent unit~(GRU) network, so that improvements are attributable to the data and not to the architecture.
This data-level approach complements the architectural priors of~\citet{DaLio2023}.\looseness=-1

To the best of our knowledge, this is the first study to systematically examine the cross-vehicle transfer of data-driven lateral-velocity estimators on production passenger vehicles, and the first to improve it by physically motivated data augmentation.
The main contributions are as follows.

\begin{itemize}
\item The first systematic input-set comparison for data-driven $\vy$ estimation on production vehicles, in which six input sets are evaluated by their error on unseen vehicles.
\item Three physically motivated augmentations of the training data, each evaluated individually across vehicles. Symmetry Data Augmentation exploits the mirror symmetry of road vehicles, and two further augmentations emulate signal delays and accelerometer offsets in the measurement chain.
\item Empirical quantification of the cross-vehicle gap in real-world test drives with three series-production passenger vehicles, together with a comparison against a model-based estimator under a zero-shot condition.
\end{itemize}

Three experiments, detailed in Section~\ref{sec:design}, support these contributions.
Experiment~1 trains nine GRUs on each of the six input sets and assesses how strongly the input set determines cross-vehicle accuracy.
Experiment~2 ablates the three augmentations individually, and Experiment~3 combines the augmentations that improve the transfer and compares the resulting GRU against a model-based estimator.
A qualitative time-series evaluation on selected maneuvers of both unseen vehicles closes the results.

The remainder of this paper is structured as follows.
Section~\ref{sec:methods} introduces the vehicles and datasets, the estimators, and the experimental design.
Section~\ref{sec:results} presents the results, which are discussed in Section~\ref{sec:discussion}.
Finally, Section~\ref{sec:conclusion} summarizes the main findings and concludes the paper.

	\begin{figure*}[t]
		\centering
		\resizebox{1.0\linewidth}{!}{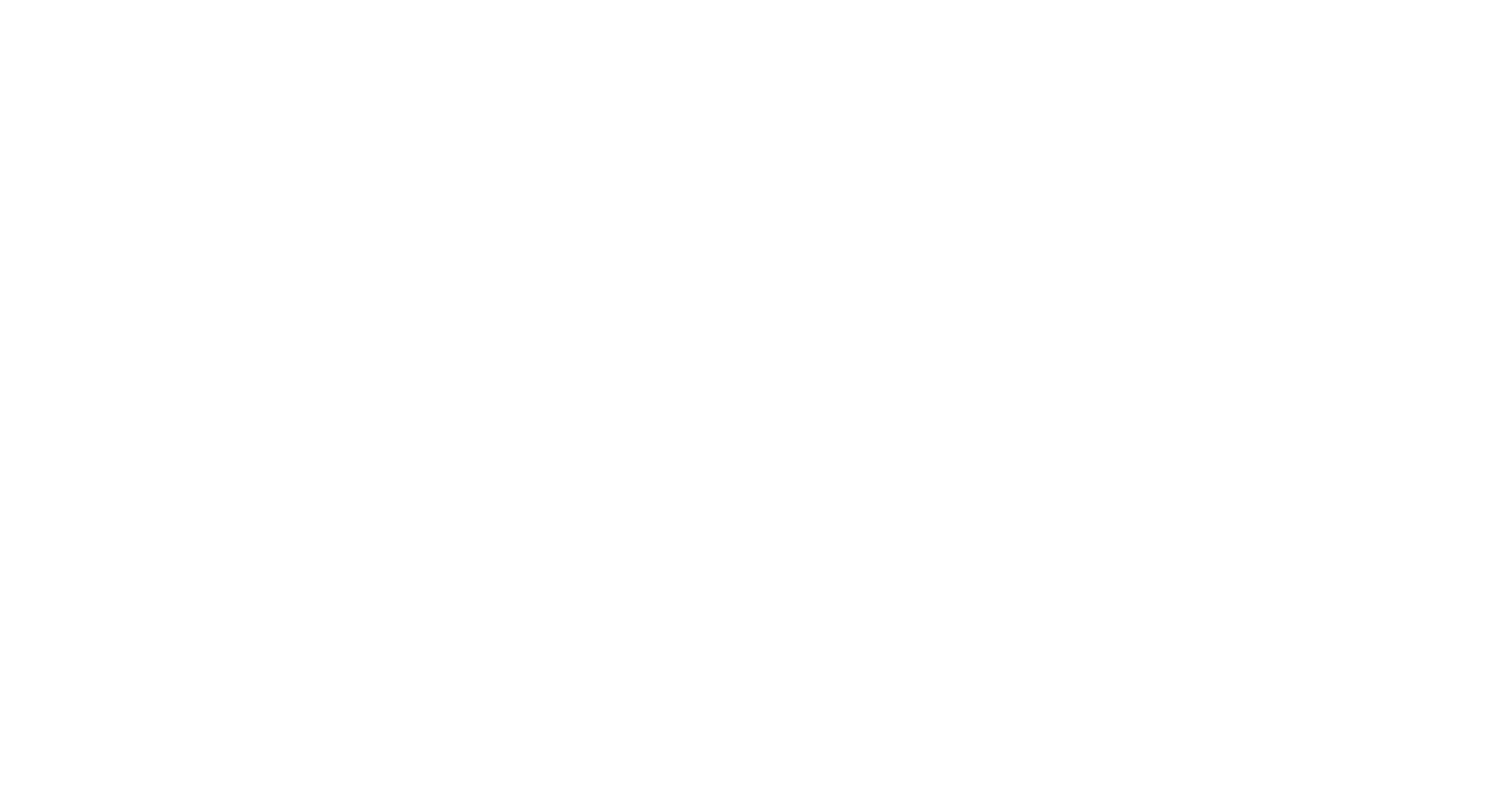}
		\caption{Overview of the study on cross-vehicle zero-shot estimation of the lateral velocity $\vy$.
(a)~A GRU estimator is trained on production-CAN data of Vehicle~A and transferred zero-shot to the structurally different, unseen Vehicle~B and Vehicle~C. All three vehicles are drawn generically for confidentiality, with illustrations generated by Google Gemini 3.1 Pro.
(b)~The estimation pipeline with the four measures. The input-set selection acts on the input signals, and the three data augmentations act on the training windows only.
(c)~Estimated $\vy$ during a cornering maneuver on the unseen Vehicle~B, where augmentation restores the tracking of the ADMA reference that the GRU without augmentation loses.}\label{fig:overview}
\end{figure*}

\section{Materials and Methods}
\label{sec:methods}
This section describes the vehicles and data, the input sets and augmentations, the two estimators, and the experimental design.
Every estimator in this study is trained only on the training data of Vehicle~A and evaluated on the data of the unseen Vehicle~B and Vehicle~C, with no signal of the unseen vehicles entering the training.
Figure~\ref{fig:overview} shows the transfer setting, the four proposed measures, and a resulting prediction on an unseen vehicle.

\subsection[Vehicles and Measurement Setup]{Vehicles and Measurement Setup}
\label{sec:data}

All data are self-collected recordings from real-world test drives with three series-production passenger vehicles.
Vehicle~A, a compact plug-in hybrid, serves as the single training vehicle.
The battery-electric Vehicle~B (SUV) and Vehicle~C (van) remain unseen during training and are evaluated zero-shot.
The make and model of all three vehicles are withheld for reasons of industrial confidentiality.
The unseen vehicles differ substantially from the training vehicle in body type, drivetrain, mass, and wheelbase (Table~\ref{tab:vehicles}).
The recordings stem from separate campaigns on different test tracks, on different days, and with different drivers, so the transfer includes realistic differences in the measurement setup and the driving data.
On all three vehicles, an Automotive Dynamic Motion Analyzer (ADMA-G-ECO+, GeneSys Elektronik GmbH), a GNSS-aided inertial system, provides the reference for $\vy$.
The unit receives real-time kinematic (RTK) correction data over a mobile network.
It is specified with root-mean-square errors of \SI{0.03}{\kilo\metre\per\hour} in velocity and \SI{0.1}{\degree} in sideslip angle.
Using its measured mounting offsets, it transforms its outputs to the center of gravity.

\begin{table}[tb]
\centering
\footnotesize
\setlength{\tabcolsep}{4pt}
\caption{Main characteristics of the three vehicles. The values of Vehicle~B and Vehicle~C are rounded.}
\label{tab:vehicles}
\begin{tabular}{@{}lccc@{}}
\toprule
 & Vehicle~A & Vehicle~B & Vehicle~C \\
\midrule
Role & training & unseen & unseen \\
Body type & compact car & SUV & van \\
Drivetrain & plug-in hybrid & battery-electric & battery-electric \\
Driven axles & front & rear & front and rear \\
Mass in \si{\kilogram} & \num{1720} & \num{2150} & \num{2650} \\
Wheelbase in \si{\meter} & \num{2.62} & \num{2.8} & \num{3.0} \\
Source & measured & data sheet & data sheet \\
\bottomrule
\end{tabular}
\end{table}

The candidate inputs of the GRU are ten signals of the production CAN bus, namely the yaw rate $\dot{\psi}$ and the longitudinal and lateral accelerations $a_x$ and $a_y$ of the inertial measurement unit (IMU), the road-wheel steering angle $\delta$, the four wheel speeds $n_\mathrm{FL}$, $n_\mathrm{FR}$, $n_\mathrm{RL}$, $n_\mathrm{RR}$ (front left, front right, rear left, rear right), and the drive and brake torques.

The preprocessing synchronizes the signals, resamples them to a common rate of \SI{50}{\hertz} with a zero-order hold, cuts them into fixed-length windows, and normalizes each~\mbox{channel}.

The data of Vehicle~A comprise \SI{22}{\minute} of driving in \num{27} maneuvers, which are divided by maneuver into training, validation, and test maneuvers (${\approx}\,80/10/10\,\%$ by samples).
The data of Vehicle~B and Vehicle~C consist only of dynamic handling maneuvers and are partitioned into validation maneuvers (\num{4.2} and \SI{3.3}{\minute}) and test maneuvers (\num{2.1} and \SI{1.6}{\minute}).
The training maneuvers of Vehicle~A serve to train all GRUs and to tune the model-based estimator, and the validation maneuvers of Vehicle~A serve the hyperparameter search (Stage~0) and give the in-domain error.
The validation maneuvers of Vehicle~C serve the selection of the input set in Experiment~1, of the augmentations in Experiment~2, and of their strengths in the preliminary sweep, and the validation maneuvers of both unseen vehicles give the out-of-domain error.
The test maneuvers of all three vehicles take no part in any selection and are reported separately in Experiment~3.
The same holds for the validation maneuvers of Vehicle~B, so that no design decision of this study rests on data of this vehicle.\looseness=-1

\begin{figure}[pos=t]
\centering
\includegraphics[width=\linewidth]{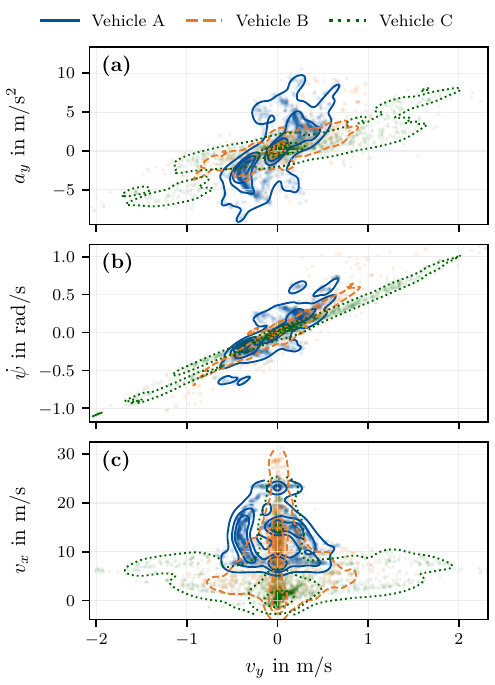}
\caption{Dataset distributions of the three vehicles over all splits: (a)~lateral acceleration $a_y$, (b)~yaw rate $\dot{\psi}$, and (c)~longitudinal velocity $v_x$, each against the lateral velocity $\vy$ (ADMA reference).
Contour lines show the sample density per vehicle (one color and line style each). The faint points are individual samples and preserve the distribution tails.
The same $\vy$ arises at different combinations of $a_y$ and $\dot{\psi}$ on the three vehicles, reflecting their different dynamics (mainly the differing masses), and the campaigns cover different $v_x$ ranges. These two differences correspond to the vehicle gap and the measurement gap.\looseness=-1}
\label{fig:dist}
\end{figure}

Figure~\ref{fig:dist} shows the dataset distributions of the three vehicles.
The vehicles reach the same $\vy$ at different combinations of $a_y$ and $\dot{\psi}$, and the campaigns cover different speed ranges.
The two observations correspond to the two sources of the transfer problem.
Cross-vehicle transfer is an instance of dataset shift, in which the estimator is trained under one distribution of signals and dynamics and deployed under another~\citep{QuinoneroCandela2009}.
Within this shift we distinguish two sources, the \emph{vehicle gap} and the \emph{measurement gap}.
The vehicle gap arises from physical parameters that differ between vehicles, such as mass, yaw inertia, center-of-gravity position, and cornering stiffness.
In the taxonomy of dataset shift, this is a change in the strength of the factors generating the dynamics~\citep{QuinoneroCandela2009}.
Consistent with this, \citet{Kontos2025} suggest that the vehicles' weight distribution drives the accuracy loss they observe across vehicles, and \citet{James2020} find that the parameters of the same longitudinal model, identified separately on two vehicles, differ by orders of magnitude.
The measurement gap arises from properties of the measurement chain, such as sensor-specific latencies, varying CAN transmission delays~\citep{Mori2024}, drifting sample rates, and individual accelerometer biases.
These properties of the measurement chain differ between vehicles and campaigns.
The measurement gap also covers the recordings themselves, which differ in the routes driven, the maneuvers covered, and the balance of left and right cornering.
Similar shifts are documented in inertial-sensing systems~\citep{Oishi2025}.

\subsection{Input-Set Variation}
\label{sec:inputsets}

To examine how the choice of input signals affects the cross-vehicle transfer, we treat the input set itself as the design variable and reduce its number of signals stepwise.
Table~\ref{tab:inputsets} lists the resulting family of six input sets.
Starting from the full ten-signal pool (Set~1), which contains the drive and brake torques used by prior neural-network estimators~\citep{Ziaukas2021}, the sets successively drop the torques (Set~2), average the wheel speeds per side (Set~3) and over all four wheels (Set~4), drop the wheel speeds entirely (Set~5), and finally reduce to the three IMU signals (Set~6).
The working hypothesis is that vehicle-specific signals help in-domain but hurt transfer, in line with the observation that feeding too many inputs ties a neural network too closely to the training dynamics~\citep{Tristano2024PINN}.
The torque signals are the clearest example, because the same torque accelerates two vehicles of different mass differently.
The goal is an input set that transfers well and is as~lean~as~possible.

Beyond transfer accuracy, a lean input set is also preferable in production.
It depends on fewer individual signals that can be degraded or lost in operation, such as the wheel speeds, which lose resolution at low velocity through quantization and latency~\citep{Schaefke2026}.
Fewer inputs also mean fewer signals to monitor and fewer plausibility checks to implement in a series-production safety concept.
We further expect a leaner input set to lower the required amount of training data and the required number of neural-network parameters, though this study does not test it directly.\looseness=-1

Sensor-light designs do exist in the literature, but they fix the input set instead of varying it.
\citet{Gonzalez2020}, for example, replace the individual wheel speeds with a preprocessed longitudinal velocity~\citep{Tanelli2006, Klomp2014}.
Every prior estimator considered here fixes one such representation (four raw wheel speeds, a preprocessed $v_x$, or none) without treating the granularity as a design~variable.

\begin{table*}[t]
\centering
\footnotesize \setlength{\tabcolsep}{4pt} \caption{The six input sets as a signal matrix, with $n$ the number of input signals.
Every set contains the IMU core $a_x$, $a_y$, $\dot\psi$, so the sets differ only in the wheel-speed resolution, the steering angle, and the drive and brake torques.
Wheel speeds enter per wheel ($n_\mathrm{FL}, n_\mathrm{FR}, n_\mathrm{RL}, n_\mathrm{RR}$), per side as left and right averages ($n_\mathrm{L}, n_\mathrm{R}$), or as a single all-wheel average ($n_\mathrm{avg}$).
The last column names the published neural-network $\vy$ estimator with the most similar input set, where (approx.) marks a close but not exact match, and \emph{none} a set not used in the literature so far.} 
\label{tab:inputsets}
\begin{tabular}{@{}cc l ccc c cc p{3.4cm}@{}}
\toprule
Set & $n$ inputs & Wheel speeds & \multicolumn{3}{c}{IMU} & Steering & \multicolumn{2}{c}{Torques} & Closest prior work \\
\cmidrule(lr){4-6}\cmidrule(lr){8-9}
    &            &              & $a_x$ & $a_y$ & $\dot\psi$ & $\delta$ & $M_\mathrm{drive}$ & $M_\mathrm{brake}$ & \\
\midrule
1 & 10 & four (per wheel) & \checkmark & \checkmark & \checkmark & \checkmark & \checkmark & \checkmark & \cite{Ziaukas2021,Srinivasan2020} (approx.) \\
2 & 8  & four (per wheel) & \checkmark & \checkmark & \checkmark & \checkmark & --         & --         & \cite{Graeber2019,Giuliacci2023,Ghosh2018,Bonfitto2020,NapolitanoDellAnnunziata2024} (approx.) \\
3 & 6  & two (per side)   & \checkmark & \checkmark & \checkmark & \checkmark & --         & --         & \emph{none} \\
4 & 5  & one (average)    & \checkmark & \checkmark & \checkmark & \checkmark & --         & --         & \cite{Kim2020,Gonzalez2020,Melzi2011,DaLio2023} (approx.) \\
5 & 4  & none             & \checkmark & \checkmark & \checkmark & \checkmark & --         & --         & \emph{none} \\
6 & 3  & none             & \checkmark & \checkmark & \checkmark & --         & --         & --         & \emph{none} \\
\bottomrule
\end{tabular}
\end{table*}

\subsection{Data Augmentation}
\label{sec:augmentation}
The three augmentations transform the training windows of Vehicle~A during training, each with its own probability $p$, so the GRU sees recorded and augmented windows in every epoch.
They change neither the GRU nor the training procedure, and the GRU is trained on the same number of batches with and without augmentation.
Algorithm~\ref{alg:aug} and the following subsections describe the three augmentations, Symmetry Data Augmentation (SDA), Channel Time Shift, and IMU Bias.
The augmentation strengths and application probabilities were chosen in a preliminary sweep over SDA probabilities from \num{0} to \num{0.7}, maximum time shifts up to \SI{5}{\milli\second} at probabilities from \num{0.2} to \num{0.8}, and bias standard deviations from \num{0} to \SI{0.5}{\meter\per\second\squared}, with five runs per setting.
Each value was set to the one that minimized the error on the validation maneuvers of Vehicle~C and then held fixed.

\subsubsection{Symmetry Data Augmentation}
\label{sec:sda}
Symmetry Data Augmentation enforces the left/right symmetry of the vehicle at the data level.
It negates the lateral signals ($a_y$, $\dot{\psi}$, $\delta$) and the target $\vy$, swaps the left and right wheel speeds ($n_\mathrm{FL} \leftrightarrow n_\mathrm{FR}$, $n_\mathrm{RL} \leftrightarrow n_\mathrm{RR}$), and leaves the longitudinal acceleration $a_x$ and the torques unchanged (Algorithm~\ref{alg:aug}, panel~a).
In the taxonomy of time-series augmentation, this is a flipping transform under an assumed symmetry~\citep{Wen2021}, which here follows from the mirror symmetry of the vehicle about its longitudinal $xz$ plane.
SDA thus introduces this symmetry as an observational bias~\citep{Karniadakis2021}, which favors a symmetric estimate without enforcing it.
In each training batch, all windows are mirrored with probability $p = 0.4$ and are otherwise left unchanged.
This reduces the left/right imbalance of the training data of Vehicle~A, in which \SI{56}{\percent} of the cornering samples ($|a_y| > \SI{1}{\meter\per\second\squared}$) are right turns, to about \SI{51}{\percent} in training.

The wheel swap requires separate left and right wheel speeds, which only some input sets provide.
The averaged wheel speed of Set~4 is invariant under mirroring, and Sets~5 and~6 do not contain any wheel speeds at all.
Therefore, SDA for these sets reduces to the sign flip of the lateral signals.
The improvements reported in Section~\ref{sec:res-ablation} thus result solely from the sign flip.
The swap is active only for Sets~1 through~3, and its isolated contribution is quantified separately in Appendix~\ref{app:sda-swap}.

\subsubsection{Channel Time Shift Augmentation}
\label{sec:timeshift}
The remaining two augmentations address the measurement chain.
The Channel Time Shift augmentation addresses time offsets between the input channels, which arise from fluctuating CAN bus latencies and sampling-time jitter between maneuvers and vehicles.
During training, each input channel is independently delayed by a random amount $\alpha_c \sim \mathcal{U}(0, \alpha_{\max})$ samples, with linear interpolation for sub-sample shifts (Algorithm~\ref{alg:aug}, panel~b).
Only the inputs are shifted, and the target $\vy$ keeps its original timing.
The neural network is thereby trained to infer the same lateral velocity from inputs whose relative timing varies slightly.
The maximum shift is $\alpha_{\max} = 0.15$ samples, or \SI{3}{\milli\second} at \SI{50}{\hertz}, and the augmentation is applied with probability $p = 0.2$ per training window.

\subsubsection{IMU Bias Augmentation}
\label{sec:imubias}
The IMU Bias augmentation emulates a constant offset of the measured accelerations.
Such an offset can be caused by the vehicle's tilt (road gradient, load), installation tolerances, and sensor drift.
During training, a constant offset $b_c \sim \mathcal{N}(0, \sigma^2)$ is drawn per window and added to the measured accelerations $a_x$ and $a_y$ before the channel normalization (Algorithm~\ref{alg:aug}, panel~c).
The offset standard deviation is $\sigma = \SI{0.3}{\meter\per\second\squared}$ for both channels, and the augmentation is applied per training window with probability $p = 0.4$.
This standard deviation equals the gravity component of a static tilt of about \SI{1.8}{\degree}.

\begin{algorithm*}[tp]
\caption{Data augmentations applied to the training data of Vehicle~A.
(a)~Symmetry Data Augmentation, (b)~Channel Time Shift, and (c)~IMU Bias, each shown with an example of its effect on a signal.
$\mm{S}$ input sequence with channels $s_c$, $\vy$ target, $p$ application probability, $\alpha_{\max}$ maximum shift in samples, $\sigma$ bias standard deviation in \si{\meter\per\second\squared} on the unnormalized acceleration signals.\looseness=-1}
\label{alg:aug}
\begin{minipage}[c]{0.53\linewidth}
\textbf{(a)}\par\vspace{2pt}
\begin{algorithmic}[1]
\Procedure{SDA}{$\mm{S},\vy,p$}
  \If{$\mathrm{rand}()<p$}
    \State $a_y,\dot\psi,\delta \gets -a_y,\,-\dot\psi,\,-\delta$\Comment{mirror inputs}
    \State $\mathrm{swap}(n_\mathrm{FL},n_\mathrm{FR})$\Comment{if present}
    \State $\mathrm{swap}(n_\mathrm{RL},n_\mathrm{RR})$\Comment{if present}
    \State $v_y \gets -v_y$\Comment{mirror target}
  \EndIf
  \State \Return $\mm{S},\vy$
\EndProcedure
\end{algorithmic}
\end{minipage}\hfill
\begin{minipage}[c]{0.43\linewidth}
\includegraphics[width=\linewidth]{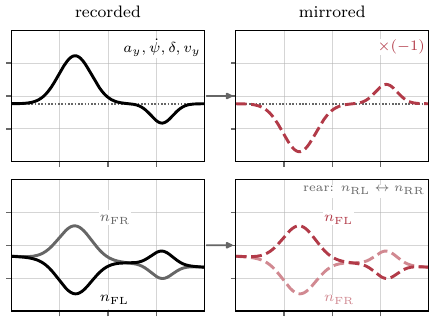}
\end{minipage}

\vspace{6pt}\hrule\vspace{6pt}
\begin{minipage}[c]{0.53\linewidth}
\textbf{(b)}\par\vspace{2pt}
\begin{algorithmic}[1]
\Procedure{ChannelTimeShift}{$\mm{S},p,\alpha_{\max}$}
  \State $\mm{S}'\gets\mm{S}$
  \If{$\mathrm{rand}()<p$}
    \For{each channel $c$}
      \State $\alpha_c \sim \mc{U}(0,\alpha_{\max})$
      \State $s'_c[t]\gets(1-\alpha_c)\,s_c[t]$\Comment{linear interp.}
      \Statex \hskip4.2em $+\,\alpha_c\,s_c[\max(0,t{-}1)]\quad\forall t$
    \EndFor
  \EndIf
  \State \Return $\mm{S}'$
\EndProcedure
\end{algorithmic}
\end{minipage}\hfill
\begin{minipage}[c]{0.43\linewidth}
\includegraphics[width=\linewidth]{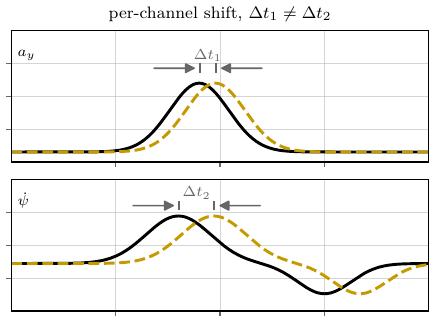}
\end{minipage}

\vspace{6pt}\hrule\vspace{6pt}
\begin{minipage}[c]{0.53\linewidth}
\textbf{(c)}\par\vspace{2pt}
\begin{algorithmic}[1]
\Procedure{IMUBias}{$\mm{S},p,\sigma$}
  \If{$\mathrm{rand}()<p$}
    \For{$c\in\{a_x,a_y\}$}\Comment{unnormalized}
      \State $b_c\sim\mc{N}(0,\sigma^2)$;\ $s_c\gets s_c+b_c$
    \EndFor
  \EndIf
  \State \Return $\mm{S}$
\EndProcedure
\end{algorithmic}
\end{minipage}\hfill
\begin{minipage}[c]{0.43\linewidth}
\includegraphics[width=\linewidth]{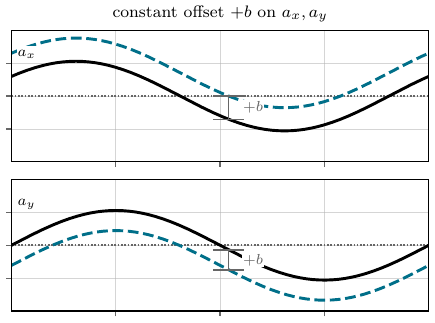}
\end{minipage}
\end{algorithm*}

\subsection{Neural Network Architecture and Training}
\label{sec:model}
Each GRU maps the signals of its input set to the lateral velocity $\vy$ over time.
It consists of a single GRU layer~\citep{Cho2014_GRU} and a linear output layer.
We have used this simple, purely data-driven architecture without physical priors before for related vehicle-state estimation~\citep{Schaefke2023}.
For training, the maneuvers are cut into windows of \numrange{150}{250} samples, with the length selected per input set (Appendix~\ref{app:hyperparams}) and a step size of one sample between consecutive windows.
Within each comparison, the architecture, its hyperparameters, and the inference path are identical across all conditions, so that the ablation can isolate the effect of the augmentations.

All GRUs are trained for \num{100} epochs with the mean absolute error (MAE) as the loss and AdamW~\citep{Loshchilov2019_AdamW} as the optimizer.
The number of epochs was determined in an exploratory preliminary study covering the range $\{50, 100, 150, 200\}$.
In an exploratory comparison of four adaptive optimizers related to Adam~\citep{Kingma2017_Adam}, namely AdamW, RAdam~\citep{Liu2020_RAdam}, Ranger~\citep{Wright2019_Ranger}, and SOAP~\citep{Vyas2024_SOAP}, the choice among them changed the cross-vehicle root-mean-square error~(RMSE) only negligibly.
The GRUs are implemented in PyTorch using the tsfast library~\citep{Weber2024_tsfast}.\looseness=-1

\subsection{Model-Based Estimator}
\label{sec:ekf}

To test whether the data-driven estimator reaches model-based accuracy on unseen vehicles, we include a classical extended Kalman filter~(EKF) on a dynamic single-track model with a linear tire model, a standard choice in the model-based estimation literature~\citep{Puscul2024, Antonov2011}.
Its state comprises the lateral velocity and the yaw rate, the steering angle is the control input, and the measured lateral acceleration and yaw rate are the two measurements.
The longitudinal velocity, computed from the wheel speeds, enters the system matrices at every step as an exogenous scheduling variable.
All inputs of the EKF thus come from the production CAN bus, as do those of the data-driven estimator.\looseness=-1

Like the data-driven estimator, the EKF is tuned exclusively on the training maneuvers of Vehicle~A.
Particle swarm optimization~\citep{Kennedy1995_PSO} tunes the two cornering stiffnesses and the four filter noise covariances by minimizing the summed per-maneuver RMSE of $\vy$ against the ADMA reference over these maneuvers, using \num{100} particles over \num{50} iterations.
The remaining physical parameters (mass, yaw inertia, and the axle distances) are measured on Vehicle~A and kept fixed.
The resulting set of parameters is then applied unchanged to Vehicle~B and Vehicle~C, although in practice the mass and wheelbase of each vehicle would be taken from its data sheet.
Neither is the EKF adapted to the unseen vehicles through its parameters nor is the GRU retrained on them, so both estimators face the same zero-shot condition.
The measured and the tuned parameters are listed in Appendix~\ref{app:hyperparams} (Table~\ref{tab:ekf-params}).\looseness=-1

\subsection{Experimental Design and Metrics} 
\label{sec:design}

The study consists of a hyperparameter optimization (Stage~0) and three experiments, namely the input-set comparison (Experiment~1), the augmentation ablations (Experiment~2), and the combined augmentation with its comparison against the EKF (Experiment~3).

Stage~0 performs hyperparameter optimization with the asynchronous successive halving algorithm (ASHA)~\citep{Li2020_ASHA} separately for each of the six input sets, with \num{1000} trials per set.
Each trial trains for at most \num{100} epochs, and ASHA stops underperforming trials early, with a grace budget of \num{10} epochs and a reduction factor of \num{4}.
The search covers five hyperparameters, namely hidden size, window length, learning rate, input dropout, and weight decay.
The number of GRU layers (one), the batch size (\num{32}), the optimizer, and the loss are kept constant across all trials.
The selected configuration per input set is listed in Appendix~\ref{app:hyperparams} (Table~\ref{tab:asha-configs}).

ASHA returns a single optimal configuration per input set, selected by the loss on the validation maneuvers of Vehicle~A.
With \num{1000} trials per set, the score of this winner is optimistic, since it is partly due to a favorable weight initialization and since it is the best of all trials on the very split it was selected on.
The selected configuration is therefore retrained from scratch nine times, so that the reported cross-vehicle differences reflect the input set and the augmentations, and not just a lucky run.
Each of these runs, referred to as a seed, starts from a different random weight initialization of the neural network.

In Experiment~1, nine GRUs with different seeds are trained on Vehicle~A for each of the six input sets (\num{54} GRUs in total) and evaluated on all three vehicles.
The input set is selected on the validation maneuvers of Vehicle~C and used in all subsequent experiments.
Experiment~2 evaluates the three augmentations individually, comparing nine seeds trained with one augmentation against nine trained without it, in-domain (Vehicle~A) and out-of-domain (Vehicle~B and Vehicle~C).
Experiment~3 combines the augmentations that improved the transfer to Vehicle~C in Experiment~2 and compares the resulting GRU against the EKF under the same zero-shot condition.
Both augmentations are applied independently during training, each with its probability from Section~\ref{sec:augmentation}.
Experiment~3 closes with a qualitative time-series evaluation on unseen maneuvers of Vehicle~B and Vehicle~C (Figure~\ref{fig:final}).

All estimators are evaluated with the RMSE and the maximum absolute error of the lateral velocity against the ADMA reference.
Both metrics are computed per maneuver after a warm-up of \SI{1}{\second}, which discards the initial settling phase of the EKF and the GRUs.
Per vehicle, the RMSE is averaged over the maneuvers and the maximum absolute error is taken over them.
The cross-vehicle error of an estimator is the mean of its per-vehicle RMSE over the two unseen vehicles.
For the GRUs, every reported value is the mean over nine seeds, and the nine-seed spread is the standard deviation over these seeds.\looseness=-1

\section{Results}
\label{sec:results}
The results follow the order of Section~\ref{sec:design}.
The ASHA-selected configuration per input set is listed in Appendix~\ref{app:hyperparams} (Table~\ref{tab:asha-configs}).
Across the six searches of \num{1000} trials each, about \SI{75}{\percent} of the trials were stopped at the grace budget of \num{10} epochs and a further \SI{18}{\percent} at the intermediate budget of \num{40} epochs.
Only \numrange{58}{86} trials per set received the full \num{100}-epoch budget.
Because ASHA promotes trials asynchronously, against the results recorded so far, this number varies per set around the synchronous expectation of $\num{1000}/4^2\approx\num{63}$.

\subsection[Input-Set Comparison (Experiment 1)]{Input-Set Comparison (Experiment 1)}
\label{sec:res-inputsets}

\begin{figure}[pos=t]
\centering
\includegraphics[width=\linewidth]{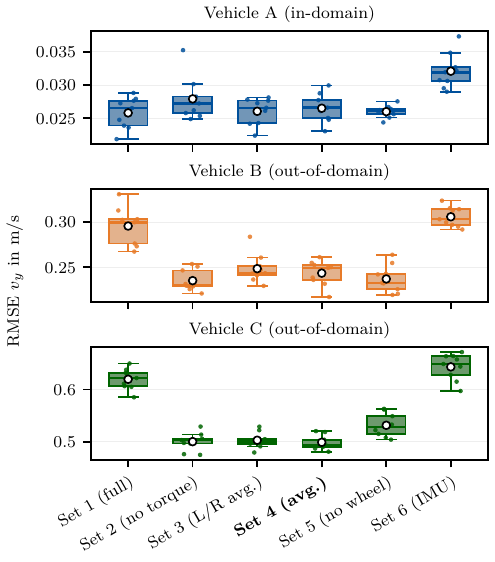}
\caption{Validation RMSE of $\vy$ across the six input sets (Experiment~1), one panel per vehicle.
Boxes span the interquartile range of the nine seeds with the median as the center line, whiskers extend to 1.5 times the interquartile range, and the open circle marks the per-set mean.
The full set (Set~1) and the IMU-only set (Set~6) degrade most on the unseen vehicles, whereas the mid-sized sets transfer best.}
\label{fig:exp1}
\end{figure}

Figure~\ref{fig:exp1} shows the validation RMSE of the six input sets of Table~\ref{tab:inputsets} per vehicle as box plots over the nine seeds.
Sets~2 to~5 reach similar cross-vehicle errors between \num{0.368} and \num{0.384}\,\si{\meter\per\second}, and their differences lie within the scatter over the nine seeds.
The full ten-signal set (Set~1, \num{0.458}\,\si{\meter\per\second}) and the IMU-only set (Set~6, \num{0.475}\,\si{\meter\per\second}) transfer worst.
Among these sets, Set~4 (five signals, \num{0.371}\,\si{\meter\per\second}) is the leanest set that still carries a speed signal, and it is also the most robust across seeds and vehicles, with the tightest nine-seed spread and the lowest worst-vehicle error.
With the lowest error on Vehicle~C, Set~4 is therefore adopted together with its Stage~0 configuration for all following experiments.

Sets~1 and~2 differ only in the two torque signals, so their comparison shows what these signals gain in-domain and what they cost across vehicles.
In-domain the two sets are equivalent on average (\num{0.026} versus \num{0.028}\,\si{\meter\per\second}, within the nine-seed spread), and the best individual neural network is trained with the torques (\num{0.022} versus \num{0.025}\,\si{\meter\per\second}).
On the unseen vehicles, the full set is clearly worse (\num{0.458} versus \num{0.368}\,\si{\meter\per\second}).

\subsection{Augmentation Ablations (Experiment 2)}
\label{sec:res-ablation}

\begin{figure}[pos=t]
\centering
\includegraphics[width=\linewidth]{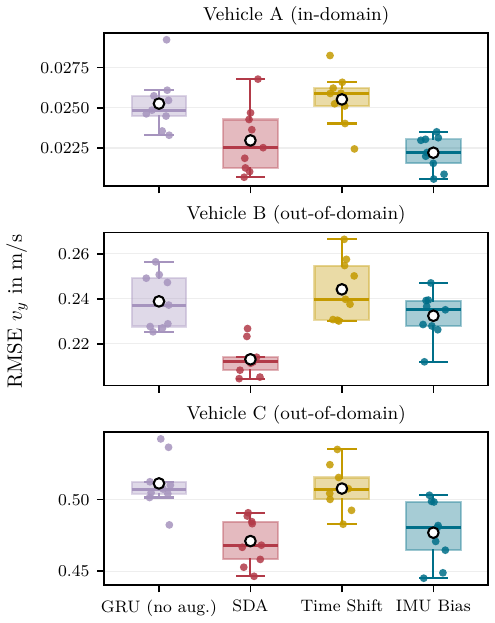}
\caption{Augmentation ablation on Set~4 (Experiment~2), one panel per vehicle, with the GRU without augmentation next to the three augmentations applied individually. Each box aggregates the nine seeds of one condition and is drawn as in Figure~\ref{fig:exp1}, with the seeds added as dots. Symmetry Data Augmentation lowers the out-of-domain error on both unseen vehicles and IMU Bias on Vehicle~C, whereas Channel Time Shift does not, which motivates its exclusion from the combined augmentation.\looseness=-1}
\label{fig:ablation}
\end{figure}

Figure~\ref{fig:ablation} compares the three augmentations individually against the GRU without augmentation on all three vehicles.
Without augmentation, Set~4 reaches an RMSE of \num{0.027}\,\si{\meter\per\second} in-domain on Vehicle~A, \num{0.244}\,\si{\meter\per\second} on Vehicle~B ($\sim$9$\times$), and \num{0.499}\,\si{\meter\per\second} on Vehicle~C ($\sim$19$\times$).
All conditions of this experiment are retrained, including the GRU without augmentation, so its errors differ marginally from these values of Experiment~1 and are the basis for every percentage reported here and in the following sections.

Symmetry Data Augmentation gives the largest out-of-domain gain, \SI{11}{\percent} on Vehicle~B and \SI{8}{\percent} on Vehicle~C, and it improves the in-domain accuracy on Vehicle~A by \SI{9}{\percent}.
IMU Bias lowers the error by \SI{3}{\percent} on Vehicle~B, within the nine-seed spread, and by \SI{7}{\percent} on Vehicle~C.
Channel Time Shift has no positive effect on cross-vehicle transfer and is therefore not considered in the next experiment.

\subsection{Combined Augmentation and Model-Based Comparison (Experiment 3)}
\label{sec:res-comparison}

\begin{figure}[pos=t]
\centering
\includegraphics[width=\linewidth]{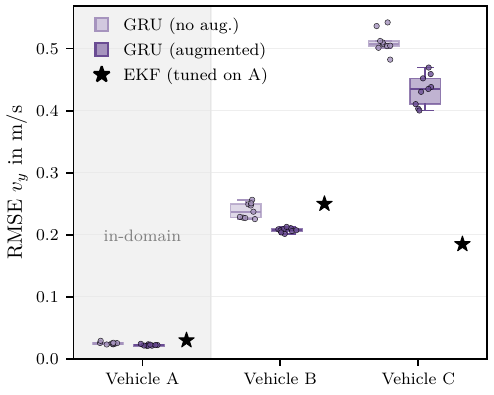}
\caption{Cross-vehicle RMSE of $\vy$, grouped per vehicle.
Boxes show the GRU without and with the combined augmentation (boxes and whiskers as in Figure~\ref{fig:exp1}, dots for the individual seeds), and the black star marks the model-based EKF tuned on Vehicle~A.
On these validation maneuvers, the augmented GRU reaches a lower error than the EKF on Vehicle~B, and the EKF reaches a lower error on Vehicle~C.}
\label{fig:comparison}
\end{figure}

Figure~\ref{fig:comparison} compares the GRU without augmentation, the same GRU trained with the combined augmentation of SDA and IMU Bias, and the EKF tuned on Vehicle~A.
Table~\ref{tab:results} lists the RMSE and the maximum absolute error of both GRUs and the EKF on the validation and the test maneuvers of all three vehicles.
Combining the two augmentations that improved the transfer to Vehicle~C in Experiment~2 reduces the out-of-domain RMSE by \SI{13}{\percent} on Vehicle~B and by \SI{15}{\percent} on Vehicle~C, and the in-domain RMSE on Vehicle~A decreases by \SI{11}{\percent}.

On the test maneuvers, the augmentation leaves the RMSE nearly unchanged on Vehicle~A and lowers it by \SI{4}{\percent} on Vehicle~B and by \SI{9}{\percent} on Vehicle~C.

\begin{table*}[t]
\centering
\footnotesize
\caption{RMSE and maximum absolute error of $\vy$ in \si{\meter\per\second} in Experiment~3 for the GRU without and with the combined augmentation and for the EKF tuned on Vehicle~A.
GRU values are means over nine seeds, with the standard deviation of the RMSE in parentheses, and the maximum error is that of the worst maneuver.
The input set and the augmentations are selected on the validation maneuvers of Vehicle~C, and the test maneuvers take no part in any~selection.
Bold marks the lowest value per vehicle and column.}
\label{tab:results}
\input{figures/tab_results.tex}
\end{table*}

The maximum absolute error decreases as well, by \SI{48}{\percent} on Vehicle~A, \SI{11}{\percent} on Vehicle~B, and \SI{32}{\percent} on Vehicle~C.
On the validation maneuvers, the standard deviation over the nine seeds decreases by about a factor of three on Vehicle~B, whereas on Vehicle~C it increases by about a third, so that the augmentation improves the accuracy on this vehicle without stabilizing the result across seeds.

On Vehicle~A, the EKF reaches an RMSE of \num{0.030}\,\si{\meter\per\second}, close to the \num{0.025}\,\si{\meter\per\second} of the GRU without augmentation, so both estimators enter the transfer with comparable in-domain accuracy.
On the validation maneuvers of Vehicle~B the augmented GRU is more accurate than the EKF (\num{0.208} versus \num{0.250}\,\si{\meter\per\second}), whereas on its test maneuver the EKF is more accurate (\num{0.125} versus \num{0.173}\,\si{\meter\per\second}).
Over the validation maneuvers, the maximum absolute error of the EKF is almost twice that of the augmented GRU, whereas on the test maneuver both are close (\num{0.81} versus \num{0.79}\,\si{\meter\per\second}).
On Vehicle~C the EKF reaches less than half the error of the augmented GRU on both splits (\num{0.185} versus \num{0.433}\,\si{\meter\per\second} on the validation maneuvers and \num{0.216} versus \num{0.572}\,\si{\meter\per\second} on the test maneuver).
The validation maneuvers of this vehicle consist of tight circular driving at sustained lateral velocities outside the range covered by the training data of Vehicle~A.

\begin{figure*}[tp]
\centering
\includegraphics[width=\linewidth]{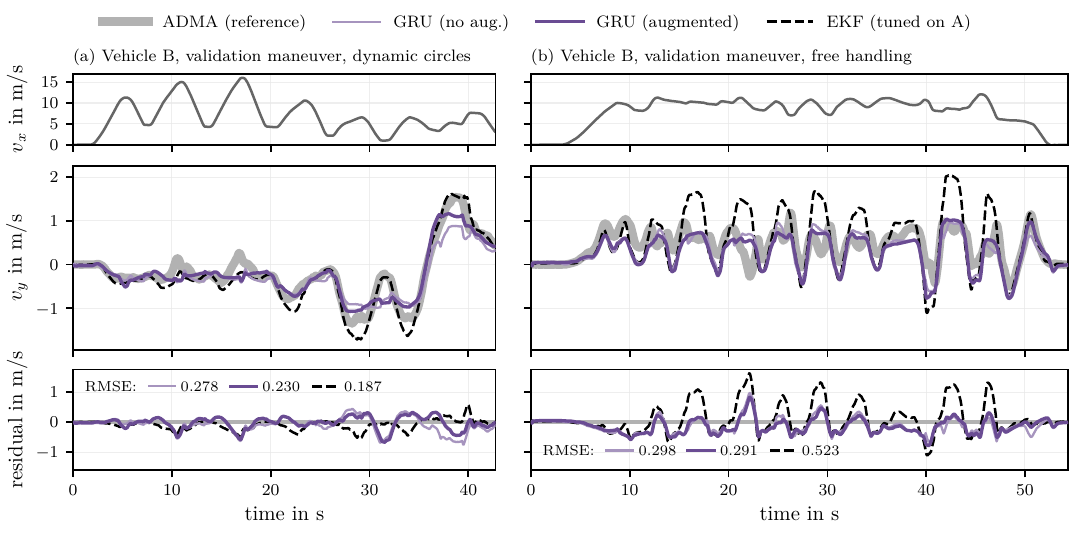}\par\medskip
\includegraphics[width=\linewidth]{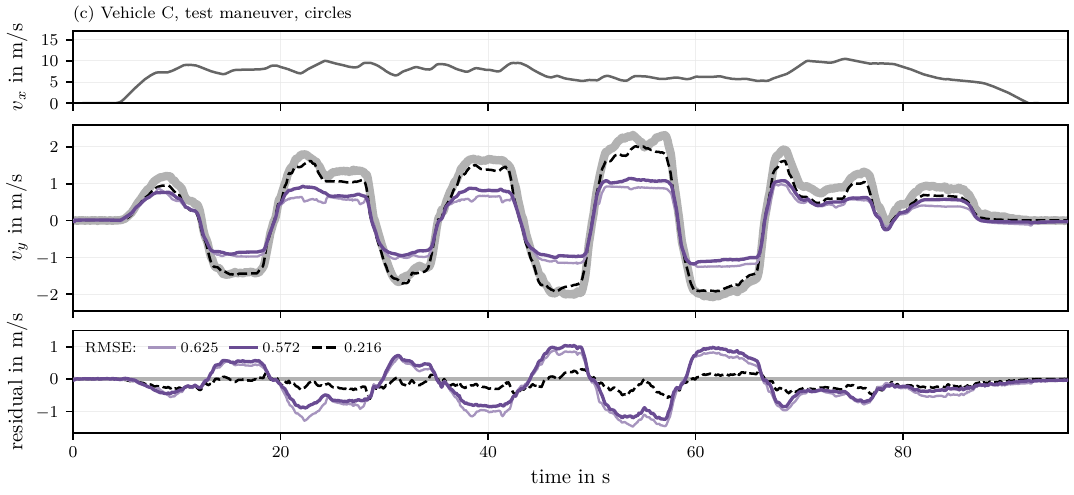}
\caption{Qualitative cross-vehicle prediction on two validation maneuvers of Vehicle~B, (a) dynamic circles and (b) free handling, and on the test maneuver of Vehicle~C, (c) circles.
Top: longitudinal velocity $v_x$ of the ADMA reference for maneuver context.
Middle: lateral velocity $\vy$ of the ADMA reference, the GRU without and with the combined augmentation, and the EKF, all trained or tuned on Vehicle~A.
Bottom: residuals with the maneuver RMSE of each estimator.\looseness=-1}
\label{fig:final}
\end{figure*}

Figure~\ref{fig:final} shows the GRU with and without the combined augmentation and the EKF against the ADMA reference on three maneuvers that take no part in any selection, two of Vehicle~B and one of Vehicle~C.
The two maneuvers of Vehicle~B show the typical gain of the augmentation and the largest deviation of the EKF, and Vehicle~C contributes its only test maneuver.
On the dynamic circles of Vehicle~B (Figure~\ref{fig:final}a), the augmentation lowers the RMSE from \num{0.278} to \num{0.230}\,\si{\meter\per\second}, mainly in the sustained cornering phases, where the GRU without augmentation underestimates the amplitude of $\vy$.
In the free handling maneuver (Figure~\ref{fig:final}b), both GRUs reach about \SI{0.29}{\meter\per\second}, whereas the EKF overshoots in most cornering phases and reaches \num{0.523}\,\si{\meter\per\second}.
On Vehicle~C (Figure~\ref{fig:final}c), both GRUs level off at about \SI{1}{\meter\per\second} in the sustained circles, while the ADMA reference reaches about \SI{2}{\meter\per\second}, beyond the range of the training data of Vehicle~A, and the EKF follows these amplitudes with an RMSE of \num{0.216}\,\si{\meter\per\second} against \num{0.572}\,\si{\meter\per\second} for the augmented GRU.\looseness=-1

\section{Discussion}
\label{sec:discussion}

The three experiments show that the cross-vehicle error of a purely data-driven estimator can be reduced without training data from the unseen vehicles, while a considerable gap to the in-domain accuracy remains on both unseen vehicles.

\subsection{Input Selection, Augmentation, and the Remaining Gap}

The input-set comparison shows that the drive and brake torques, the signals tied to the training vehicle, impair transfer without a reliable in-domain gain (Section~\ref{sec:res-inputsets}).
The most plausible mechanism is a scaling mismatch, because the same torque acts on a different mass through a different drivetrain ratio and therefore implies a different reaction on each vehicle, while the GRU has learned the reaction of the training vehicle.
The wheel-speed granularity, varied here for the first time, turns out to be largely immaterial, since side-wise and all-wheel aggregation lose nothing relative to the four individual speeds.
What transfers is therefore the vehicle speed itself and not the individual wheel speeds, whereas removing the speed information entirely increases the error on Vehicle~C.

Even with this input set, the cross-vehicle gap without augmentation remains large, about $9\times$ the in-domain error on Vehicle~B and about $19\times$ on Vehicle~C, which differs most from Vehicle~A in mass and body type and whose validation maneuvers also exceed the lateral-velocity range of the training data.
Symmetry Data Augmentation reduces an imbalance between left and right cornering that a single recording campaign does not average out, and it improves the transfer to both unseen vehicles whereas its in-domain gain remains on the order of the nine-seed spread.
Of the two augmentations that address the measurement chain, only IMU Bias transfers, adding robustness against accelerometer offsets.
Channel Time Shift brings no cross-vehicle gain in the form tested here, as random per-channel delays of up to \SI{3}{\milli\second}.\looseness=-1

The expectation that model-based estimators transfer better than data-driven ones is common in the literature but rarely tested~\citep{Graeber2019, Liu2021}.
On Vehicle~B the augmented GRU reaches an accuracy comparable to that of the EKF and shows none of the collapse of a black-box neural network reported by~\citet{DaLio2023}, whereas on Vehicle~C the EKF remains clearly more accurate.
On Vehicle~C the lateral velocities exceed the range of the training data of Vehicle~A, which the single-track model covers through its structure while the GRU has to extrapolate.
The EKF is not free of the transfer problem either, since its parameters are vehicle-specific and would ordinarily be re-identified per vehicle.
Both families retain a substantial cross-vehicle gap.
The largest absolute errors of the GRU occur on Vehicle~C and those of the EKF on Vehicle~B, and the EKF has the larger RMSE on Vehicle~B on the validation maneuvers but on Vehicle~C on the test maneuver.
The error of the EKF therefore depends on the maneuvers as well as on the vehicle, and with one recording campaign per vehicle the two effects cannot be separated.

For a safety-relevant estimator the worst-case error matters at least as much as the mean, and the augmentation lowers the maximum absolute error on all three vehicles, on the validation and the test maneuvers (Section~\ref{sec:res-comparison}).
The augmentations themselves are a few lines of training-time code, need no training data from the target vehicle and no additional sensors, and add no inference-time overhead.
At about \num{24000} parameters, the GRU is of a size that suits an automotive electronic control unit.
The symmetry argument rests on the geometry of the vehicle and the remaining arguments on the measurement chain, so neither is specific to the estimated state and the same augmentations apply to other vehicle-dynamics estimation targets.

\subsection{Limitations}
The input set, the augmentations, and their strengths are selected on the validation maneuvers of Vehicle~C.
In practice this is a one-time design decision made on a few vehicles with reference measurements and then transferred to the fleet, and confirming this choice on further vehicles would remove the residual selection dependence.
All results rest on a single training vehicle and a single architecture, and the augmentations, which are architecture-agnostic, are not demonstrated here on other backbones.
The EKF is evaluated with the parameters of Vehicle~A only.

Progress on cross-vehicle estimation is further limited by the data available in the field, because the three campaigns used here are, as is typical, inhomogeneous in scope and length, and no public multi-vehicle dataset combines reference measurements with production signals.
Public benchmarks of this kind would make such studies comparable across research groups.

\section{Conclusion}
\label{sec:conclusion}
In this paper, we evaluated the extent to which a data-driven lateral-velocity estimator trained on a single production vehicle transfers to structurally different production vehicles and investigated which data-level measures can improve this transfer.
A GRU estimator was trained on production-CAN signals of Vehicle~A and evaluated zero-shot on Vehicle~B and Vehicle~C against the ADMA reference, alongside a model-based EKF tuned on the same training data under the identical zero-shot condition.\looseness=-1

The systematic input-set comparison shows that the input set is an important design choice for transfer.
Among the sets that are practically equivalent in-domain, within \SI{8}{\percent} of one another, the cross-vehicle error differs by up to \SI{25}{\percent}.
What remains is a transfer-robust set of five production-CAN signals, the three IMU signals, the steering angle, and an averaged wheel speed.
Symmetry Data Augmentation exploits the lateral symmetry of vehicle dynamics, and the IMU Bias augmentation targets the accelerometer-bias component of the measurement gap, whereas Channel Time Shift shows no cross-vehicle benefit and is dropped.
Without augmentation, the out-of-domain error exceeds the in-domain error by an order of magnitude on both unseen vehicles.
The augmented GRU, trained with Symmetry Data Augmentation and IMU Bias, has a \SI{13}{\percent} lower out-of-domain error than the GRU without augmentation on the validation maneuvers of Vehicle~B and a \SI{15}{\percent} lower one on those of Vehicle~C.
On Vehicle~B the augmented GRU reaches an accuracy comparable to that of the EKF, and on Vehicle~C the EKF remains clearly more accurate.

The results show that the cross-vehicle gap in data-driven vehicle-state estimation can be narrowed using simple and low-cost measures that require no training data from the target vehicle.
Future work could extend the training to multiple training vehicles to expose the neural network to a wider range of vehicle parameters and thereby further reduce the cross-vehicle error.
Simulation could supply such a range at low cost if it also models the production measurement chain, and the transfer from simulated to real vehicles would then need its own validation.
These data-level measures could also be combined with architectural priors~\citep{DaLio2023}, which address the problem from a complementary perspective.
A further logical step is to apply the proposed data augmentations to other vehicle-dynamics estimation problems.
Online adaptation on the target vehicle represents another promising direction.

\printcredits

\section*{Declaration of competing interest}
A.~Vagapov and A.~Trabelsi are employees of IAV GmbH, which funded this work.
C.~Schweers was employed by IAV GmbH when this work was performed.
The other authors report no competing interests.

\section*{Funding}
This work was supported by IAV GmbH.

\section*{Data availability statement}
The data of this study are not publicly available due to confidentiality agreements with the industrial partner.

\section*{Declaration of generative AI and AI-assisted technologies in the manuscript preparation process}
During the preparation of this work the authors used Claude Opus 5.5 (Anthropic) to improve the readability and language of the manuscript and to assist with code, and Gemini 3.1 Pro (Google) to generate the generic vehicle illustrations in Figure~\ref{fig:overview}(a).
After using these tools, the authors reviewed and edited the content as needed and take full responsibility for the content of the published article.

\bibliographystyle{unsrtnat}
\providecommand{\doi}{}\renewcommand{\doi}[1]{doi:\,\href{https://doi.org/#1}{{\urlstyle{rm}\nolinkurl{#1}}}}
\bibliography{literatur}

\appendix
\counterwithin{table}{section}
\counterwithin{figure}{section}
\section{Estimator and Optimization Details}
\label{app:hyperparams}

Table~\ref{tab:ekf-params} lists the measured and the tuned parameters of the EKF (Section~\ref{sec:ekf}).
Table~\ref{tab:asha-configs} lists the ASHA-selected GRU configuration per input set (Stage~0).
Across all six sets the neural network is a single-layer GRU trained for \num{100} epochs with batch size \num{32}, window length varying per set.
Only the values that differ between sets are shown.

\begin{table}[pos=!h]
\centering
\caption{EKF parameters, measured on Vehicle~A or tuned on its training maneuvers.
All parameters are applied unchanged to Vehicle~B and Vehicle~C.
The noise covariances are listed as standard deviations, with $\mm{Q} = \operatorname{diag}(q_{v_y}^2, q_{\dot\psi}^2)$ and $\mm{R} = \operatorname{diag}(r_{a_y}^2, r_{\dot\psi}^2)$.}
\label{tab:ekf-params}
\begin{tabular}{l l l}
\toprule
Group & Parameter & Value \\
\midrule
Measured & mass $m$                & \SI{1720}{\kilogram} \\
                 & yaw inertia $I_z$       & \SI{2066}{\kilogram\meter\squared} \\
                 & front axle distance $l_\mathrm{f}$ & \SI{1.16}{\meter} \\
                 & rear axle distance $l_\mathrm{r}$  & \SI{1.46}{\meter} \\
\midrule
Tuned      & front cornering stiffness $C_\mathrm{f}$ & \SI{140510}{\newton\per\radian} \\
                 & rear cornering stiffness $C_\mathrm{r}$  & \SI{177693}{\newton\per\radian} \\
                 & process noise $q_{v_y},\,q_{\dot\psi}$ & \num{0.129},\ \num{0.076} \\
                 & measurement noise $r_{a_y},\,r_{\dot\psi}$ & \num{4.42},\ \num{0.954} \\
\bottomrule
\end{tabular}
\end{table}

\begin{table}[pos=!h]
\centering
\footnotesize
\caption{ASHA-selected GRU configuration per input set (Stage~0), trained on Vehicle~A.
Fixed across all sets: single-layer GRU, AdamW optimizer, \num{100} epochs, MAE loss, batch size \num{32}, recurrent dropout and auxiliary losses off.}
\label{tab:asha-configs}
\begin{tabular*}{\columnwidth}{@{\extracolsep{\fill}}c r l l l r@{}}
\toprule
Set & Hidden & Learning & Input & Weight & Window \\
    & size   & rate     & dropout & decay & (samples) \\
\midrule
1 &  86 & \num{7.2e-4} & \num{2.0e-2} & \num{3.5e-5} & 200 \\
2 & 124 & \num{3.8e-4} & \num{4.3e-3} & \num{3.1e-5} & 200 \\
3 & 124 & \num{9.6e-4} & \num{3.7e-2} & \num{4.8e-4} & 200 \\
4 &  86 & \num{3.3e-4} & \num{3.8e-3} & \num{7.8e-5} & 250 \\
5 &  95 & \num{4.5e-4} & \num{4.1e-4} & \num{1.1e-3} & 150 \\
6 & 104 & \num{8.1e-4} & \num{3.1e-2} & \num{4.3e-4} & 250 \\
\bottomrule
\end{tabular*}
\end{table}

\section{Component-Wise Ablation of the Symmetry Data Augmentation}
\label{app:sda-swap}

\begin{figure}[pos=!t]
\centering
\includegraphics[width=\linewidth]{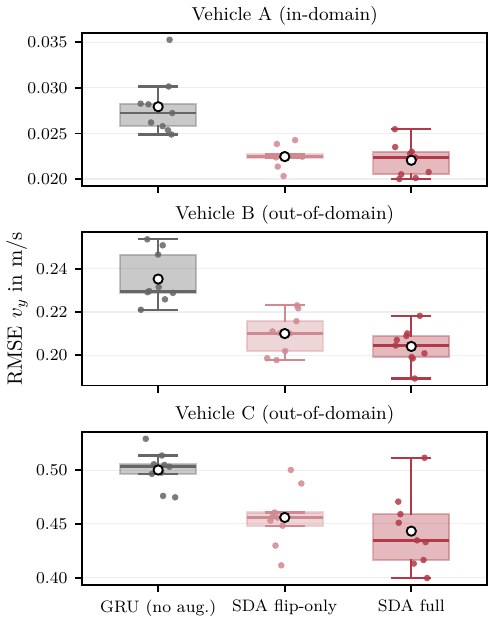}
\caption{Component-wise ablation of Symmetry Data Augmentation on input Set~2, which carries the four individual wheel speeds so that the left/right swap is active, with one panel per vehicle. The three conditions per panel are the GRU without augmentation from Experiment~1, the sign flip alone (\textsc{flip-only}), and the sign flip combined with the wheel swap (\textsc{full}), and each box aggregates the nine seeds with the individual seeds as dots. Most of the gain is due to the sign flip, while the wheel swap adds a small further cross-vehicle reduction.}
\label{fig:sda-ablation}
\end{figure}

Symmetry Data Augmentation (Section~\ref{sec:sda}) combines a sign flip of the lateral signals with a left/right swap of the wheel speeds. 
On the selected input set the averaged wheel speed is invariant under mirroring, so the swap has no effect there and the reported gains stem from the sign flip alone.
To isolate the swap, the augmentation is repeated on Set~2, which carries the four individual wheel speeds, with the sign flip only (\textsc{flip-only}) and with both operations (\textsc{full}).
Figure~\ref{fig:sda-ablation} shows that the sign flip accounts for most of the improvement, reducing the out-of-domain RMSE by \SI{11}{\percent} on Vehicle~B and by \SI{9}{\percent} on Vehicle~C.
The swap adds a further \SI{3}{\percent} on both unseen vehicles, within the nine-seed spread, and leaves the in-domain error on Vehicle~A unchanged at \num{0.022}\,\si{\meter\per\second}.

\end{document}

%% file: figures/overview.pdf_tex
\begingroup%
  \makeatletter%
  \providecommand\color[2][]{%
    \errmessage{(Inkscape) Color is used for the text in Inkscape, but the package 'color.sty' is not loaded}%
    \renewcommand\color[2][]{}%
  }%
  \providecommand\transparent[1]{%
    \errmessage{(Inkscape) Transparency is used (non-zero) for the text in Inkscape, but the package 'transparent.sty' is not loaded}%
    \renewcommand\transparent[1]{}%
  }%
  \providecommand\rotatebox[2]{#2}%
  \newcommand*\fsize{\dimexpr\f@size pt\relax}%
  \newcommand*\lineheight[1]{\fontsize{\fsize}{#1\fsize}\selectfont}%
  \ifx\svgwidth\undefined%
    \setlength{\unitlength}{862.5bp}%
    \ifx\svgscale\undefined%
      \relax%
    \else%
      \setlength{\unitlength}{\unitlength * \real{\svgscale}}%
    \fi%
  \else%
    \setlength{\unitlength}{\svgwidth}%
  \fi%
  \global\let\svgwidth\undefined%
  \global\let\svgscale\undefined%
  \makeatother%
  \begin{picture}(1,0.52869565)%
    \lineheight{1}%
    \setlength\tabcolsep{0pt}%
    \put(0,0){\includegraphics[width=\unitlength,page=1]{overview.pdf}}%
    \put(0.02086957,0.49043478){\color[rgb]{0.10196078,0.10196078,0.10196078}\makebox(0,0)[lt]{\lineheight{1.25}\smash{\begin{tabular}[t]{l}\textbf{\textbf{(a)}}\end{tabular}}}}%
    \put(0,0){\includegraphics[width=\unitlength,page=2]{overview.pdf}}%
    \put(0.2173913,0.32782609){\color[rgb]{0,0.25098039,0.48627451}\makebox(0,0)[t]{\lineheight{1.25}\smash{\begin{tabular}[t]{c}\textbf{\textbf{Vehicle A}}\end{tabular}}}}%
    \put(0.2173913,0.30391304){\color[rgb]{0.10196078,0.10196078,0.10196078}\makebox(0,0)[t]{\lineheight{1.25}\smash{\begin{tabular}[t]{c}training vehicle\end{tabular}}}}%
    \put(0,0){\includegraphics[width=\unitlength,page=3]{overview.pdf}}%
    \put(0.45391304,0.40713043){\color[rgb]{0.10196078,0.10196078,0.10196078}\makebox(0,0)[t]{\lineheight{1.25}\smash{\begin{tabular}[t]{c}\textbf{\textbf{zero-shot transfer}}\end{tabular}}}}%
    \put(0,0){\includegraphics[width=\unitlength,page=4]{overview.pdf}}%
    \put(0.6626087,0.32782609){\color[rgb]{0.7254902,0.38431373,0.12941176}\makebox(0,0)[t]{\lineheight{1.25}\smash{\begin{tabular}[t]{c}\textbf{\textbf{Vehicle B}}\end{tabular}}}}%
    \put(0,0){\includegraphics[width=\unitlength,page=5]{overview.pdf}}%
    \put(0.87826087,0.32782609){\color[rgb]{0,0.31372549,0}\makebox(0,0)[t]{\lineheight{1.25}\smash{\begin{tabular}[t]{c}\textbf{\textbf{Vehicle C}}\end{tabular}}}}%
    \put(0.77043478,0.30391304){\color[rgb]{0.10196078,0.10196078,0.10196078}\makebox(0,0)[t]{\lineheight{1.25}\smash{\begin{tabular}[t]{c}unseen vehicles\end{tabular}}}}%
    \put(0.02086957,0.25869565){\color[rgb]{0.10196078,0.10196078,0.10196078}\makebox(0,0)[lt]{\lineheight{1.25}\smash{\begin{tabular}[t]{l}\textbf{\textbf{(b)}}\end{tabular}}}}%
    \put(0,0){\includegraphics[width=\unitlength,page=6]{overview.pdf}}%
    \put(0.11043478,0.22565217){\color[rgb]{0.10196078,0.10196078,0.10196078}\makebox(0,0)[t]{\lineheight{1.25}\smash{\begin{tabular}[t]{c}\textbf{\textbf{production-CAN inputs}}\end{tabular}}}}%
    \put(0.11043478,0.21086957){\color[rgb]{0.4,0.4,0.4}\makebox(0,0)[t]{\lineheight{1.25}\smash{\begin{tabular}[t]{c}Vehicle A training data\end{tabular}}}}%
    \put(0.04,0.19086957){\color[rgb]{0.10196078,0.10196078,0.10196078}\makebox(0,0)[lt]{\lineheight{1.25}\smash{\begin{tabular}[t]{l}$\cdot$ $\dot\psi$ yaw rate\end{tabular}}}}%
    \put(0.04,0.17608696){\color[rgb]{0.10196078,0.10196078,0.10196078}\makebox(0,0)[lt]{\lineheight{1.25}\smash{\begin{tabular}[t]{l}$\cdot$ $a_x,\, a_y$ acceleration\end{tabular}}}}%
    \put(0.04,0.16130435){\color[rgb]{0.10196078,0.10196078,0.10196078}\makebox(0,0)[lt]{\lineheight{1.25}\smash{\begin{tabular}[t]{l}$\cdot$ $\delta$ steering angle\end{tabular}}}}%
    \put(0.04,0.14652174){\color[rgb]{0.10196078,0.10196078,0.10196078}\makebox(0,0)[lt]{\lineheight{1.25}\smash{\begin{tabular}[t]{l}$\cdot$ 4$\times$ wheel speed\end{tabular}}}}%
    \put(0.04,0.13173913){\color[rgb]{0.10196078,0.10196078,0.10196078}\makebox(0,0)[lt]{\lineheight{1.25}\smash{\begin{tabular}[t]{l}$\cdot$ drive, brake torque\end{tabular}}}}%
    \put(0,0){\includegraphics[width=\unitlength,page=7]{overview.pdf}}%
    \put(0.23913043,0.18086957){\color[rgb]{0.10196078,0.10196078,0.10196078}\makebox(0,0)[t]{\lineheight{1.25}\smash{\begin{tabular}[t]{c}\textbf{\textbf{window}}\end{tabular}}}}%
    \put(0,0){\includegraphics[width=\unitlength,page=8]{overview.pdf}}%
    \put(0.32695652,0.18086957){\color[rgb]{0.10196078,0.10196078,0.10196078}\makebox(0,0)[t]{\lineheight{1.25}\smash{\begin{tabular}[t]{c}\textbf{\textbf{augment}}\end{tabular}}}}%
    \put(0,0){\includegraphics[width=\unitlength,page=9]{overview.pdf}}%
    \put(0.41304348,0.18086957){\color[rgb]{0.10196078,0.10196078,0.10196078}\makebox(0,0)[t]{\lineheight{1.25}\smash{\begin{tabular}[t]{c}\textbf{\textbf{GRU}}\end{tabular}}}}%
    \put(0,0){\includegraphics[width=\unitlength,page=10]{overview.pdf}}%
    \put(0.4773913,0.18086957){\color[rgb]{0.10196078,0.10196078,0.10196078}\makebox(0,0)[t]{\lineheight{1.25}\smash{\begin{tabular}[t]{c}\textbf{\textbf{$v_y$}}\end{tabular}}}}%
    \put(0,0){\includegraphics[width=\unitlength,page=11]{overview.pdf}}%
    \put(0.11043478,0.06782609){\color[rgb]{0.10196078,0.10196078,0.10196078}\makebox(0,0)[t]{\lineheight{1.25}\smash{\begin{tabular}[t]{c}\textbf{\textbf{input}}\end{tabular}}}}%
    \put(0.11043478,0.05391304){\color[rgb]{0.10196078,0.10196078,0.10196078}\makebox(0,0)[t]{\lineheight{1.25}\smash{\begin{tabular}[t]{c}\textbf{\textbf{selection}}\end{tabular}}}}%
    \put(0,0){\includegraphics[width=\unitlength,page=12]{overview.pdf}}%
    \put(0.20869565,0.09173913){\color[rgb]{0.10196078,0.10196078,0.10196078}\makebox(0,0)[lt]{\lineheight{1.25}\smash{\begin{tabular}[t]{l}\textbf{\textbf{data augmentations (training only)}}\end{tabular}}}}%
    \put(0,0){\includegraphics[width=\unitlength,page=13]{overview.pdf}}%
    \put(0.26086957,0.06782609){\color[rgb]{0.10196078,0.10196078,0.10196078}\makebox(0,0)[t]{\lineheight{1.25}\smash{\begin{tabular}[t]{c}\textbf{\textbf{Symmetry Aug.}}\end{tabular}}}}%
    \put(0.26086957,0.05391304){\color[rgb]{0.10196078,0.10196078,0.10196078}\makebox(0,0)[t]{\lineheight{1.25}\smash{\begin{tabular}[t]{c}\textbf{\textbf{(SDA)}}\end{tabular}}}}%
    \put(0,0){\includegraphics[width=\unitlength,page=14]{overview.pdf}}%
    \put(0.36565217,0.06782609){\color[rgb]{0.10196078,0.10196078,0.10196078}\makebox(0,0)[t]{\lineheight{1.25}\smash{\begin{tabular}[t]{c}\textbf{\textbf{Channel}}\end{tabular}}}}%
    \put(0.36565217,0.05391304){\color[rgb]{0.10196078,0.10196078,0.10196078}\makebox(0,0)[t]{\lineheight{1.25}\smash{\begin{tabular}[t]{c}\textbf{\textbf{Time Shift}}\end{tabular}}}}%
    \put(0,0){\includegraphics[width=\unitlength,page=15]{overview.pdf}}%
    \put(0.45652174,0.06782609){\color[rgb]{0.10196078,0.10196078,0.10196078}\makebox(0,0)[t]{\lineheight{1.25}\smash{\begin{tabular}[t]{c}\textbf{\textbf{IMU}}\end{tabular}}}}%
    \put(0.45652174,0.05391304){\color[rgb]{0.10196078,0.10196078,0.10196078}\makebox(0,0)[t]{\lineheight{1.25}\smash{\begin{tabular}[t]{c}\textbf{\textbf{Bias}}\end{tabular}}}}%
    \put(0,0){\includegraphics[width=\unitlength,page=16]{overview.pdf}}%
    \put(0.53478261,0.25869565){\color[rgb]{0.10196078,0.10196078,0.10196078}\makebox(0,0)[lt]{\lineheight{1.25}\smash{\begin{tabular}[t]{l}\textbf{\textbf{(c)}}\end{tabular}}}}%
    \put(0,0){\includegraphics[width=\unitlength,page=17]{overview.pdf}}%
  \end{picture}%
\endgroup%

%% file: figures/tab_results.tex
\begin{tabular}{l l r r r r}
\toprule
 & & \multicolumn{2}{c}{Validation maneuvers} & \multicolumn{2}{c}{Test maneuvers} \\
\cmidrule(lr){3-4} \cmidrule(lr){5-6}
Vehicle & Estimator & RMSE & Max.\ error & RMSE & Max.\ error \\
\midrule
Vehicle~A & GRU (no aug.) & 0.025 (0.002) & 0.21 & \textbf{0.015} (0.002) & 0.10 \\
 & GRU (augmented) & \textbf{0.022} (0.001) & \textbf{0.11} & 0.016 (0.003) & \textbf{0.08} \\
 & EKF & 0.030 & 0.12 & 0.035 & 0.09 \\
\midrule
Vehicle~B & GRU (no aug.) & 0.239 (0.011) & 0.97 & 0.180 (0.008) & 0.89 \\
 & GRU (augmented) & \textbf{0.208} (0.003) & \textbf{0.86} & 0.173 (0.009) & \textbf{0.79} \\
 & EKF & 0.250 & 1.63 & \textbf{0.125} & 0.81 \\
\midrule
Vehicle~C & GRU (no aug.) & 0.511 (0.017) & 1.46 & 0.625 (0.026) & 1.48 \\
 & GRU (augmented) & 0.433 (0.023) & 1.00 & 0.572 (0.029) & 1.25 \\
 & EKF & \textbf{0.185} & \textbf{0.63} & \textbf{0.216} & \textbf{0.61} \\
\bottomrule
\end{tabular}

%% file: main.bbl
\begin{thebibliography}{44}
\providecommand{\natexlab}[1]{#1}
\providecommand{\url}[1]{\texttt{#1}}
\expandafter\ifx\csname urlstyle\endcsname\relax
  \providecommand{\doi}[1]{doi: #1}\else
  \providecommand{\doi}{doi: \begingroup \urlstyle{rm}\Url}\fi

\bibitem[Jin et~al.(2019)Jin, Yin, and Chen]{Jin2019}
Xianjian Jin, Guodong Yin, and Nan Chen.
\newblock Advanced estimation techniques for vehicle system dynamic state: A
  survey.
\newblock \emph{Sensors}, 19\penalty0 (19):\penalty0 4289, 2019.
\newblock \doi{10.3390/s19194289}.

\bibitem[Puscul et~al.(2024)Puscul, Lex, Vignati, and Shao]{Puscul2024}
D.~Puscul, C.~Lex, M.~Vignati, and L.~Shao.
\newblock A literature survey on sideslip angle estimation using vehicle
  dynamics based methods.
\newblock \emph{IEEE Access}, 12:\penalty0 70263--70277, 2024.
\newblock \doi{10.1109/ACCESS.2024.3402429}.

\bibitem[Giuliacci et~al.(2024)Giuliacci, Ballesio, Fainello, Mair, and
  King]{Giuliacci2023}
T.~Giuliacci, S.~Ballesio, M.~Fainello, U.~Mair, and J.~King.
\newblock Recurrent neural network model for on-board estimation of the
  side-slip angle in a four-wheel drive and steering vehicle.
\newblock \emph{{SAE} International Journal of Passenger Vehicle Systems},
  17\penalty0 (1):\penalty0 37--48, 2024.
\newblock \doi{10.4271/15-17-01-0003}.

\bibitem[Srinivasan et~al.(2020)Srinivasan, Sa, Zyner, Reijgwart, Valls, and
  Siegwart]{Srinivasan2020}
Sirish Srinivasan, Inkyu Sa, Alex Zyner, Victor Reijgwart, Miguel~I. Valls, and
  Roland Siegwart.
\newblock End-to-end velocity estimation for autonomous racing.
\newblock \emph{IEEE Robotics and Automation Letters}, 5\penalty0 (4):\penalty0
  6869--6875, 2020.
\newblock \doi{10.1109/LRA.2020.3016929}.

\bibitem[Napolitano~Dell'Annunziata et~al.(2024)Napolitano~Dell'Annunziata,
  Ruffini, Stefanelli, Adiletta, Fichera, and
  Timpone]{NapolitanoDellAnnunziata2024}
G.~Napolitano~Dell'Annunziata, M.~Ruffini, R.~Stefanelli, G.~Adiletta,
  G.~Fichera, and F.~Timpone.
\newblock Four-wheeled vehicle sideslip angle estimation: A machine
  learning-based technique for real-time virtual sensor development.
\newblock \emph{Applied Sciences}, 14\penalty0 (3):\penalty0 1036, 2024.
\newblock \doi{10.3390/app14031036}.

\bibitem[James et~al.(2020)James, Anderson, and Da~Lio]{James2020}
S.~S. James, S.~R. Anderson, and M.~Da~Lio.
\newblock Longitudinal vehicle dynamics: A comparison of physical and
  data-driven models under large-scale real-world driving conditions.
\newblock \emph{IEEE Access}, 8:\penalty0 73714--73729, 2020.
\newblock \doi{10.1109/ACCESS.2020.2988592}.

\bibitem[Ziaukas et~al.(2021)Ziaukas, Busch, and Wielitzka]{Ziaukas2021}
Z.~Ziaukas, A.~Busch, and M.~Wielitzka.
\newblock Estimation of vehicle side-slip angle at varying road friction
  coefficients using a recurrent artificial neural network.
\newblock In \emph{2021 IEEE Conference on Control Technology and Applications
  ({CCTA})}, pages 986--991, 2021.
\newblock \doi{10.1109/CCTA48906.2021.9658710}.

\bibitem[Kim et~al.(2020)Kim, Min, Kim, and Huh]{Kim2020}
D.~Kim, K.~Min, H.~Kim, and K.~Huh.
\newblock Vehicle sideslip angle estimation using deep ensemble-based adaptive
  {Kalman} filter.
\newblock \emph{Mechanical Systems and Signal Processing}, 144:\penalty0
  106862, 2020.
\newblock \doi{10.1016/j.ymssp.2020.106862}.

\bibitem[Gr{\"a}ber et~al.(2019)Gr{\"a}ber, Lupberger, Unterreiner, and
  Schramm]{Graeber2019}
T.~Gr{\"a}ber, S.~Lupberger, M.~Unterreiner, and D.~Schramm.
\newblock A hybrid approach to side-slip angle estimation with recurrent neural
  networks and kinematic vehicle models.
\newblock \emph{IEEE Transactions on Intelligent Vehicles}, 4\penalty0
  (1):\penalty0 39--47, 2019.
\newblock \doi{10.1109/TIV.2018.2886687}.

\bibitem[Liu et~al.(2020{\natexlab{a}})Liu, Wang, Zhang, and Walker]{Liu2021}
J.~Liu, Z.~Wang, L.~Zhang, and P.~Walker.
\newblock Sideslip angle estimation of ground vehicles: a comparative study.
\newblock \emph{IET Control Theory \& Applications}, 14\penalty0 (20):\penalty0
  3490--3505, 2020{\natexlab{a}}.
\newblock \doi{10.1049/iet-cta.2020.0516}.

\bibitem[Melzi and Sabbioni(2011)]{Melzi2011}
S.~Melzi and E.~Sabbioni.
\newblock On the vehicle sideslip angle estimation through neural networks:
  Numerical and experimental results.
\newblock \emph{Mechanical Systems and Signal Processing}, 25\penalty0
  (6):\penalty0 2005--2019, 2011.
\newblock \doi{10.1016/j.ymssp.2010.10.015}.

\bibitem[Ghosh et~al.(2018)Ghosh, Tonoli, and Amati]{Ghosh2018}
J.~Ghosh, A.~Tonoli, and N.~Amati.
\newblock A deep learning based virtual sensor for vehicle sideslip angle
  estimation: Experimental results.
\newblock {SAE} Technical Paper 2018-01-1089, {SAE} International, 2018.

\bibitem[Bonfitto et~al.(2020)Bonfitto, Feraco, Tonoli, and
  Amati]{Bonfitto2020}
A.~Bonfitto, S.~Feraco, A.~Tonoli, and N.~Amati.
\newblock Combined regression and classification artificial neural networks for
  sideslip angle estimation and road condition identification.
\newblock \emph{Vehicle System Dynamics}, 58\penalty0 (11):\penalty0
  1766--1787, 2020.
\newblock \doi{10.1080/00423114.2019.1645860}.

\bibitem[Prieto~Gonz{\'a}lez et~al.(2020)Prieto~Gonz{\'a}lez, Sanz~S{\'a}nchez,
  Garcia-Guzman, Boada, and Boada]{Gonzalez2020}
L.~Prieto~Gonz{\'a}lez, S.~Sanz~S{\'a}nchez, J.~Garcia-Guzman, M.~J.~L. Boada,
  and B.~L. Boada.
\newblock Simultaneous estimation of vehicle roll and sideslip angles through a
  deep learning approach.
\newblock \emph{Sensors}, 20\penalty0 (13):\penalty0 3679, 2020.
\newblock \doi{10.3390/s20133679}.

\bibitem[Hermansdorfer et~al.(2021)Hermansdorfer, Trauth, Betz, and
  Lienkamp]{Hermansdorfer2020}
L.~Hermansdorfer, R.~Trauth, J.~Betz, and M.~Lienkamp.
\newblock End-to-end neural network for vehicle dynamics modeling.
\newblock In \emph{2020 6th IEEE Congress on Information Science and Technology
  ({CiSt})}, pages 407--412, 2021.
\newblock \doi{10.1109/CiSt49399.2021.9357196}.

\bibitem[Bertipaglia et~al.(2024)Bertipaglia, Alirezaei, Happee, and
  Shyrokau]{Bertipaglia2024}
A.~Bertipaglia, M.~Alirezaei, R.~Happee, and B.~Shyrokau.
\newblock An unscented {Kalman} filter-informed neural network for vehicle
  sideslip angle estimation.
\newblock \emph{IEEE Transactions on Vehicular Technology}, 73\penalty0
  (9):\penalty0 12731--12746, 2024.
\newblock \doi{10.1109/TVT.2024.3389493}.

\bibitem[Chen et~al.(2024)Chen, Yu, Wang, Zhou, and Liu]{Chen2024}
J.~Chen, C.~Yu, Y.~Wang, Z.~Zhou, and Z.~Liu.
\newblock Hybrid modeling for vehicle lateral dynamics via {AGRU} with a
  dual-attention mechanism under limited data.
\newblock \emph{Control Engineering Practice}, 151:\penalty0 106015, 2024.
\newblock \doi{10.1016/j.conengprac.2024.106015}.

\bibitem[Hwang et~al.(2024)Hwang, Seong, Han, and Oh]{Hwang2024}
G.-Y. Hwang, M.-S. Seong, C.~H. Han, and J.-S. Oh.
\newblock Hybrid feature selection approach to finding optimal feature subsets
  for vehicle lateral velocity estimation.
\newblock \emph{IEEE Access}, 12:\penalty0 120543--120552, 2024.
\newblock \doi{10.1109/ACCESS.2024.3450968}.

\bibitem[Tristano et~al.(2024)Tristano, Lenzo, Saxton, Xu, and
  Zhang]{Tristano2024PINN}
M.~Tristano, B.~Lenzo, H.~Saxton, X.~Xu, and X.~Zhang.
\newblock Physics-infused neural network-driven investigation of vehicle
  sideslip angle.
\newblock In \emph{Advances in Dynamics of Vehicles on Roads and Tracks {III}
  ({IAVSD} 2023)}, Lecture Notes in Mechanical Engineering, pages 358--365.
  Springer, 2024.
\newblock \doi{10.1007/978-3-031-66968-2_35}.

\bibitem[Kalyanasundaram et~al.(2025)Kalyanasundaram, Chandra~Sekaran,
  St{\"a}uber, Lange, Utschick, and Botsch]{Kalyanasundaram2025}
A.~Kalyanasundaram, K.~Chandra~Sekaran, P.~St{\"a}uber, M.~Lange, W.~Utschick,
  and M.~Botsch.
\newblock Uncertainty-aware hybrid machine learning in virtual sensors for
  vehicle sideslip angle estimation.
\newblock In \emph{2025 IEEE Intelligent Vehicles Symposium ({IV})}, pages
  1440--1447, 2025.
\newblock \doi{10.1109/IV64158.2025.11097574}.

\bibitem[Da~Lio et~al.(2023)Da~Lio, Piccinini, and Biral]{DaLio2023}
M.~Da~Lio, M.~Piccinini, and F.~Biral.
\newblock Robust and sample-efficient estimation of vehicle lateral velocity
  using neural networks with explainable structure informed by kinematic
  principles.
\newblock \emph{IEEE Transactions on Intelligent Transportation Systems},
  24\penalty0 (12):\penalty0 13670--13684, 2023.
\newblock \doi{10.1109/TITS.2023.3303776}.

\bibitem[Ayyad et~al.(2021)Ayyad, Prohm, Gr{\"a}ber, Unterreiner, and
  Hilbert]{Ayyad2021}
A.~Ayyad, C.~Prohm, T.~Gr{\"a}ber, M.~Unterreiner, and M.~Hilbert.
\newblock Uncertainty estimation for neural time series with an application to
  sideslip angle estimation.
\newblock \emph{{SAE} International Journal of Connected and Automated
  Vehicles}, 4\penalty0 (3):\penalty0 247--259, 2021.
\newblock \doi{10.4271/12-04-03-0020}.

\bibitem[Kontos et~al.(2025)Kontos, B{\'o}dis, and Vathy-Fogarassy]{Kontos2025}
J.~Kontos, L.~B{\'o}dis, and {\'A}.~Vathy-Fogarassy.
\newblock Sensitivity analysis of {Long Short-Term Memory}-based neural network
  model for vehicle yaw rate prediction.
\newblock \emph{Sensors}, 25\penalty0 (5):\penalty0 1363, 2025.
\newblock \doi{10.3390/s25051363}.

\bibitem[Wen et~al.(2021)Wen, Sun, Yang, Song, Gao, Wang, and Xu]{Wen2021}
Q.~Wen, L.~Sun, F.~Yang, X.~Song, J.~Gao, X.~Wang, and H.~Xu.
\newblock Time series data augmentation for deep learning: A survey.
\newblock In \emph{Proceedings of the 30th International Joint Conference on
  Artificial Intelligence ({IJCAI})}, pages 4653--4660, 2021.
\newblock \doi{10.24963/ijcai.2021/631}.

\bibitem[Str{\"o}bel et~al.(2024)Str{\"o}bel, Mau, Puchta, and
  Fleischer]{Stroebel2024}
R.~Str{\"o}bel, M.~Mau, A.~Puchta, and J.~Fleischer.
\newblock Improving time series regression model accuracy via systematic
  training dataset augmentation and sampling.
\newblock \emph{Machine Learning and Knowledge Extraction}, 6\penalty0
  (2):\penalty0 1072--1086, 2024.
\newblock \doi{10.3390/make6020049}.

\bibitem[Karniadakis et~al.(2021)Karniadakis, Kevrekidis, Lu, Perdikaris, Wang,
  and Yang]{Karniadakis2021}
G.~E. Karniadakis, I.~G. Kevrekidis, L.~Lu, P.~Perdikaris, S.~Wang, and
  L.~Yang.
\newblock Physics-informed machine learning.
\newblock \emph{Nature Reviews Physics}, 3:\penalty0 422--440, 2021.
\newblock \doi{10.1038/s42254-021-00314-5}.

\bibitem[Li et~al.(2025)Li, Tao, Song, Li, Ye, Jin, He, Wei, and Chen]{Li2025}
L.~Li, X.~Tao, H.~Song, X.~Li, Z.~Ye, Y.~Jin, Q.~He, S.~Wei, and W.~Chen.
\newblock Symmetry-based data augmentation method for deep learning-based
  structural damage identification.
\newblock \emph{Infrastructures}, 10\penalty0 (6):\penalty0 145, 2025.
\newblock \doi{10.3390/infrastructures10060145}.

\bibitem[Qui{\~n}onero-Candela et~al.(2009)Qui{\~n}onero-Candela, Sugiyama,
  Schwaighofer, and Lawrence]{QuinoneroCandela2009}
Joaquin Qui{\~n}onero-Candela, Masashi Sugiyama, Anton Schwaighofer, and
  Neil~D. Lawrence, editors.
\newblock \emph{Dataset Shift in Machine Learning}.
\newblock Neural Information Processing. The {MIT} Press, Cambridge, MA, 2009.
\newblock ISBN 9780262170055.

\bibitem[Mori et~al.(2025)Mori, Kamekawa, Fujieda, and Mizuno]{Mori2024}
Daiki Mori, Masahito Kamekawa, Nobushige Fujieda, and Yoshihiro Mizuno.
\newblock Robust vehicle state estimation from urban driving to the limits of
  handling.
\newblock \emph{IEEE Transactions on Intelligent Transportation Systems},
  26\penalty0 (1):\penalty0 892--905, 2025.
\newblock \doi{10.1109/TITS.2024.3487117}.

\bibitem[Oishi et~al.(2026)Oishi, Birch, Roggen, and Lago]{Oishi2025}
Nobuyuki Oishi, Philip Birch, Daniel Roggen, and Paula Lago.
\newblock Physically plausible data augmentations for wearable {IMU}-based
  human activity recognition using physics simulation.
\newblock \emph{IEEE Sensors Journal}, 26\penalty0 (7):\penalty0 10445--10457,
  2026.
\newblock \doi{10.1109/JSEN.2026.3661047}.

\bibitem[Sch{\"a}fke et~al.(2026)Sch{\"a}fke, Weber, Vagapov, Schweers, Seel,
  and Ehlers]{Schaefke2026}
H.~Sch{\"a}fke, D.~O.~M. Weber, A.~Vagapov, C.~Schweers, T.~Seel, and S.~F.~G.
  Ehlers.
\newblock Neural network-based virtual wheel-speed sensor for enhanced
  low-velocity state estimation.
\newblock In \emph{{IFAC-PapersOnLine} (2026 {IFAC} World Congress, Busan,
  Republic of Korea)}, 2026.
\newblock Accepted, to appear; preprint arXiv:2605.12230.

\bibitem[Tanelli et~al.(2006)Tanelli, Savaresi, and Cantoni]{Tanelli2006}
Mara Tanelli, Sergio~M. Savaresi, and Carlo Cantoni.
\newblock Longitudinal vehicle speed estimation for traction and braking
  control systems.
\newblock In \emph{2006 IEEE International Conference on Control Applications},
  pages 2790--2795, 2006.
\newblock \doi{10.1109/CCA.2006.286032}.

\bibitem[Klomp et~al.(2014)Klomp, Gao, and Bruzelius]{Klomp2014}
Matthijs Klomp, Yunlong Gao, and Fredrik Bruzelius.
\newblock Longitudinal velocity and road slope estimation in hybrid electric
  vehicles employing early detection of excessive wheel slip.
\newblock \emph{Vehicle System Dynamics}, 52\penalty0 (SUPPL. 1):\penalty0
  172--188, 2014.
\newblock \doi{10.1080/00423114.2014.887737}.

\bibitem[Cho et~al.(2014)Cho, van Merri{\"e}nboer, G{\"u}l{\c c}ehre, Bahdanau,
  Bougares, Schwenk, and Bengio]{Cho2014_GRU}
Kyunghyun Cho, Bart van Merri{\"e}nboer, {\c C}a{\u g}lar G{\"u}l{\c c}ehre,
  Dzmitry Bahdanau, Fethi Bougares, Holger Schwenk, and Yoshua Bengio.
\newblock Learning phrase representations using {RNN} encoder--decoder for
  statistical machine translation.
\newblock In \emph{Proceedings of the 2014 Conference on Empirical Methods in
  Natural Language Processing ({EMNLP})}, pages 1724--1734, 2014.
\newblock \doi{10.3115/v1/D14-1179}.

\bibitem[Schäfke et~al.(2023)Schäfke, Lampe, and Kortmann]{Schaefke2023}
Hendrik Schäfke, Nicolas Lampe, and Karl-Philipp Kortmann.
\newblock Transformer neural networks for maximum friction coefficient
  estimation of tire-road contact using onboard vehicle sensors.
\newblock In \emph{2023 62nd IEEE Conference on Decision and Control (CDC)},
  pages 5331--5338, 2023.
\newblock \doi{10.1109/CDC49753.2023.10384175}.

\bibitem[Loshchilov and Hutter(2019)]{Loshchilov2019_AdamW}
Ilya Loshchilov and Frank Hutter.
\newblock Decoupled weight decay regularization.
\newblock In \emph{7th International Conference on Learning Representations
  ({ICLR})}, 2019.
\newblock arXiv:1711.05101.

\bibitem[Kingma and Ba(2015)]{Kingma2017_Adam}
Diederik~P. Kingma and Jimmy Ba.
\newblock {Adam}: A method for stochastic optimization.
\newblock In \emph{3rd International Conference on Learning Representations
  ({ICLR})}, 2015.
\newblock arXiv:1412.6980.

\bibitem[Liu et~al.(2020{\natexlab{b}})Liu, Jiang, He, Chen, Liu, Gao, and
  Han]{Liu2020_RAdam}
Liyuan Liu, Haoming Jiang, Pengcheng He, Weizhu Chen, Xiaodong Liu, Jianfeng
  Gao, and Jiawei Han.
\newblock On the variance of the adaptive learning rate and beyond.
\newblock In \emph{8th International Conference on Learning Representations
  ({ICLR})}, 2020{\natexlab{b}}.
\newblock arXiv:1908.03265.

\bibitem[Wright(2019)]{Wright2019_Ranger}
Less Wright.
\newblock Ranger -- a synergistic optimizer.
\newblock GitHub repository, 2019.
\newblock URL
  \url{https://github.com/lessw2020/Ranger-Deep-Learning-Optimizer}.

\bibitem[Vyas et~al.(2025)Vyas, Morwani, Zhao, Shapira, Brandfonbrener, Janson,
  and Kakade]{Vyas2024_SOAP}
Nikhil Vyas, Depen Morwani, Rosie Zhao, Itai Shapira, David Brandfonbrener,
  Lucas Janson, and Sham Kakade.
\newblock {SOAP}: Improving and stabilizing {Shampoo} using {Adam} for language
  modeling.
\newblock In \emph{International Conference on Learning Representations
  ({ICLR})}, 2025.
\newblock URL \url{https://openreview.net/forum?id=IDxZhXrpNf}.

\bibitem[Weber(2024)]{Weber2024_tsfast}
Daniel O.~M. Weber.
\newblock tsfast -- a deep learning library for time series analysis and system
  identification.
\newblock GitHub repository, 2024.
\newblock URL \url{https://github.com/daniel-om-weber/tsfast}.

\bibitem[Antonov et~al.(2011)Antonov, Fehn, and Kugi]{Antonov2011}
S.~Antonov, A.~Fehn, and A.~Kugi.
\newblock Unscented {K}alman filter for vehicle state estimation.
\newblock \emph{Vehicle System Dynamics}, 49\penalty0 (9):\penalty0 1497--1520,
  2011.
\newblock \doi{10.1080/00423114.2010.527994}.

\bibitem[Kennedy and Eberhart(1995)]{Kennedy1995_PSO}
James Kennedy and Russell Eberhart.
\newblock Particle swarm optimization.
\newblock In \emph{Proceedings of ICNN'95 - International Conference on Neural
  Networks}, volume~4, pages 1942--1948, 1995.
\newblock \doi{10.1109/ICNN.1995.488968}.

\bibitem[Li et~al.(2020)Li, Jamieson, Rostamizadeh, Gonina, Ben-Tzur, Hardt,
  Recht, and Talwalkar]{Li2020_ASHA}
Liam Li, Kevin Jamieson, Afshin Rostamizadeh, Ekaterina Gonina, Jonathan
  Ben-Tzur, Moritz Hardt, Benjamin Recht, and Ameet Talwalkar.
\newblock A system for massively parallel hyperparameter tuning.
\newblock In \emph{Proceedings of Machine Learning and Systems ({MLSys})},
  volume~2, pages 230--246, 2020.

\end{thebibliography}
